# Correlative X-ray and electron tomography for scale-bridging, quantitative analysis of complex, hierarchical particle systems

Alexander Götz [a], Fabian Lutter [b], Dennis Simon Possart [b,c], Daniel Augsburger [b], Usman Arslan [b], Sabrina Pechmann [b], Carmen Rubach [a], Moritz Buwen [a], Umair Sultan [d], Alexander Kichigin [a], Johannes Böhmer [a], Nora Vorlaufer [e], Peter Suter [f], Tor Hildebrand [f], Matthias Thommes [g], Peter Felfer [e], Nicolas Vogel [d], Katharina Breininger [c], Silke Christiansen [b], Benjamin Apeleo Zubiri [a,*] and Erdmann Spiecker [a]

[a] Institute of Micro- and Nanostructure Research (IMN) & Center for Nanoanalysis and Electron Microscopy (CENEM), Friedrich-Alexander-Universität (FAU) Erlangen-Nürnberg, IZNF, Cauerstraße 3, Erlangen, Germany

[b] Fraunhofer-Institute for Ceramic Technologies and Systems – Correlative Microscopy and Materials Data, Äußere Nürnberger Straße 62, Forchheim, Germany

[c] Professur für Informatik (Pattern Recognition), Center for AI and Data Science (CAIDAS), Julius-Maximilians-Universität Würzburg, John-Skilton-Str. 4a, Würzburg, Germany

[d] Institute of Particle Technology, Friedrich-Alexander-Universität (FAU) Erlangen-Nürnberg, Cauerstraße 4, Erlangen, Germany

[e] Institute for General Materials Properties MSEI, Friedrich-Alexander-Universität (FAU) Erlangen-Nürnberg, Martensstraße 5, Erlangen, Germany

[f] Lucid Concepts AG, Zürich, Switzerland

[g] Institute of Separation Science and Technology, Friedrich-Alexander-Universität (FAU) Erlangen-Nürnberg, Egerlandstraße 3, Erlangen, Germany

* Corresponding author: benjamin.apeleo.zubiri@fau.de

## Abstract

This study presents a comprehensive workflow for investigating particulate materials through combined 360° electron tomography (ET), nano-computed X-ray tomography (nanoCT), and micro-computed X-ray tomography (microCT), alongside a versatile sample preparation routine. The workflow enables the investigation of size, morphology, and pore systems across multiple scales, from individual particles to large hierarchical structures. A customized tapered sample shape is fabricated using focused ion beam milling with the aim to optimize each imaging technique's field of view, facilitating high-resolution analysis of small volumes containing single particles, while also allowing for large-scale studies of thousands of particles for statistical relevance. By correlating data from same locations in different imaging modalities, the approach enhances the precision of quantitative analyses. The study highlights the importance of cross-scale, correlative three-dimensional microscopy for a comprehensive understanding of complex hierarchical materials. Precise data registration, segmentation using machine learning, and multimodal imaging techniques are crucial for unlocking insights into process-structure-property relationships and thus to optimize functional, hierarchical materials.

**1. Introduction**

Functional devices and materials such as battery-[1-3] and electrolyzer[4] electrodes, stationary phase materials[5] or catalyst systems[6] feature complex hierarchical structures across multiple length scales, requiring cross-scale correlative three-dimensional (3D) microscopy for detailed visualization and thus comprehensive understanding[7-11]. Moreover, optimizing complex functional materials and devices requires correlating scale-bridging techniques to link structural features with functionality.[12, 13] A versatile preparative, analytical, and data workflow is crucial for accelerating materials optimization, enabling a feedback loop between synthesis, characterization, and data assessment.

Many of the above-mentioned systems, can exist in the form of particulate materials, where parameters such as morphology, arrangement, composition and porosity govern their functional properties. Particles can exhibit intrinsic internal pore networks. In addition, in an agglomerated form or when packed into a pellet, column or react, additional interparticle pore space is created from their packing structure. When these different pore spaces are combined, hierarchical pore systems emerge, which can be tailored to provide enhanced properties with respect to transport, reaction kinetics, or dynamic adsorption.[3, 5, 14] The assessment of particle and pore statistics such as particle and pore size, interconnectivity, tortuosity or enclosed/open porosity is key to characterize and subsequently optimize such materials. Single particles, their agglomerated forms as functional structure, and the combined intra- and interparticle pore space often extend over several length scales. Internal pores can range from the micro- (< 2 nm) to the mesoporous (2 – 50 nm) regime up to larger macropores (> 50 nm), whereas the interparticle pores are usually larger macropores.[14] Single particles can have a size from only a few nm up to dozens of µm, and their agglomerates and packing structures are typically of macroscopic dimensions.[5] The difficulty is the complete assessment of all necessary, function-determining features, which cannot be performed with only one 3D characterization technique.

Micro-computed X-ray tomography (microCT) allows non-destructive studies of complex materials and device architectures.[15-17] When combined with nano-computed X-ray tomography (nanoCT), 3D focused ion beam (FIB) scanning electron microcopy (SEM) tomography (3D-FIB-SEM), and electron tomography (ET), it enables thorough scale-bridging 3D analysis of materials and devices down to nm scale.[18, 19] ET and 3D-FIB-SEM offer detailed structural and chemical insights from the atomic[20-22] to the µm scale[23-25], while microCT bridges the µm to mm scale[3, 15]. In recent years, nanoCT has extended CT resolution to < 50 nm, even in laboratory-based instruments, providing decent overlap for correlative studies with ET.[26-28] As depicted in Figure 1a, tomography techniques face an inherent trade-off between accessible volume and resolution, requiring strategies that combine different methods to leverage their respective strengths.[9, 11, 18] High-resolution information of nm-sized features and micro- and mesopores can be accessed using transmission electron microscopy (TEM) and ET, but these techniques are limited to the investigation of sample volumes typically below (1 µm)³ (cf. Figure 1a-c).[26] On the other hand, larger macropores and especially the interparticle pore space of larger particle agglomerates or devices can be investigated using 3D-FIB-SEM, nanoCT and microCT, which enable statistically relevant 3D measurements in relatively short acquisition times, but are limited in resolution (cf. Figure 1a,b,d,e).[5] Stitching approaches allow covering larger field of views while maintaining spatial resolution by acquiring several tomographic tilt series at high resolution (typically 4 to 8) of adjacent regions of interest to cover the complete sample.[29, 30] This procedure, however, is time consuming and still limited regarding the maximum possible resolution (i.e. limited by the applied imaging technique) and specimen size (about 8-10 times the volume of a single tomography with same imaging conditions).[29] Hence, there is a need to develop scale-bridging tomography techniques by combining different microscopy techniques in correlative workflows.

A key consideration for such thorough tomography analyses, whether performed at high resolution or on larger volumes, sample preparation is key. For ET of a bulk specimen or a device, smaller pieces of the specimen such as a thin lamella or pillar with an electron-transparent thickness of typically 50 - 500 nm needs to be extracted using laser ablation and/or FIB milling and lift out.[28, 31, 32] For 360° ET, nanoCT, microCT investigation of single particles and agglomerates from powder or in solution, samples can be prepared by directly dispersing particles onto a thin tomography tip[33], controlled stamping transfer in an SEM to obtain freestanding particles on a tomography tip plateau preventing shadowing from any substrate[26, 34], or by attaching nanoparticles to wires which are connected to a tomography tip[35]. An alternative to investigate a larger amount of particles from powder is to use embedding techniques for 360° ET[36] and nano/microCT[37, 38]. The latter technique involves mixing of the particles with resin and additional carbon black particles as spacers to prevent agglomeration and thus enhance the imaging quality, extruding the mixture into tube with an inner diameter matching the field of view (FOV) of the imaging technique, and finally curing the resin to obtain a solid specimen.[37, 38] In that way, thousands of single particles can be investigated in 3D in only one sample.[36, 39] Truly correlative and scale-bridging analyses require even more sophisticated sample preparation, enabling the 3D investigation of different features sizes and sample volumes in only one specimen. A common strategy is to start with a larger specimen, which is site-specifically milled down after each investigation to fit the FOV of the subsequent imaging modality with higher resolution.[28, 40-42]

Allocating identical regions of interest (ROI) in reconstructions from multimodal imaging techniques for subsequent volume registration and merging is a challenging task.[8, 43] Aligning the sample to a customized coordinate system can facilitate navigation and ROI detection between different imaging modalities. In 2D, navigation solutions for correlative microscopy exist in life science[44] and materials science applications [45, 46].

Registration of different 3D datasets can be performed either based on fiducial markers or marker-free. Examples are the usage of fluorescent markers for correlative fluorescence microscopy, atomic force microscopy and electron microscopy/tomography[47] or utilizing both fiducial markers and feature landmarks depending on sample and feature size and applied imaging technique for correlative microCT, SEM, scanning TEM (STEM) and 3D-FIB-SEM[43]. In a multitude of correlative 3D studies in materials science, key features for registration were identified in different 3D reconstructions, e.g., using microCT, nanoCT and STEM energy-dispersive X-ray spectroscopy (EDXS) tomography[8], nanoCT, SEM-EDXS, 360° ET and TEM[28], nanoCT and FIB-SEM-EDXS[42], microCT, nanoCT and FIB-SEM-EDXS[39], and in biology using, e.g., microCT, LM, TEM[48].

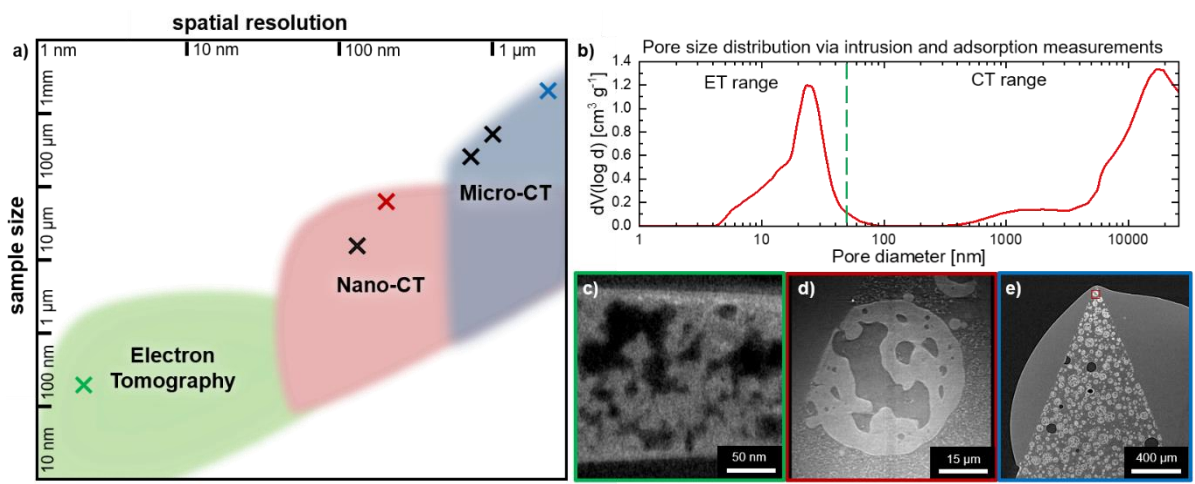

*Figure 1: (a) Overview over typical sample sizes (i.e. accessible regions of interest) and spatial resolutions for electron tomography (ET), nano-computed X-ray tomography (nanoCT) and micro-computed X-ray tomography (microCT), with crosses indicating the actual samples sizes and resolution of the different investigated reconstructions in this study (see Figure 5 and Table 2). (b) Exemplary pore size distribution of a similiar porous silica particle sample covering both meso- and macropores determined by nitrogen adsorption and mercury intrusion (dashed green line indicates the limit of 50 nm between the meso- and macropore regime). Virtual slices through reconstructed silica particles from this study with hierarchical pore structure using (c) 360° ET, (d) nanoCT, and (e) microCT (see Figures 5 and 8).*

Automated image segmentation approaches allow for fast and reproducible analysis of data, with both classical analysis pipelines that aim to exploit prior knowledge about the structures of interest, and data-driven approaches such as deep learning. The latter has received considerable attention in the last decade due to its versatility and performance even for complex structures.[49-51] Automated segmentation algorithms offer valuable insights across various technological fields, such as materials science and biomedical imaging, where precise and efficient segmentation is crucial for analyzing complex structures and extracting meaningful quantitative information.[52-54] Likewise, in the field of nanotechnology, deep learning-based segmentation workflows have emerged for identifying, separating, and analyzing high density clusters of nanoparticles across various imaging modalities.[55, 56] These methods have demonstrated their effectiveness in both ion- and electron-based microscopy, as well as in 3D tomography, e.g., nanoCT, under varying conditions, including differences in resolution and field of view.[55-57] This enables researchers to extract key particle characteristics at scale, such as size, volume, morphology, spatial distribution, and inter- and intraparticular pore structures[55, 56, 58], which are essential for understanding the resultant materials properties. Still, these approaches require a substantial amount of annotated image data for model development, where all desired structures to be segmented need to be outlined. This is typically performed manually and comes at high annotation costs. Therefore, there is a high interest in finding suitable (semi-)automated annotation strategies.

One of the biggest benefits of the above-mentioned 3D imaging techniques is that they allow for local and direct analysis of pore and particle arrangements on different scales. Even though a precise analysis of an at the same time statistically-relevant sample volume is feasible through combination of different techniques, the comparison to and validation through alternative and complementary techniques for pore and particle characterization, such as analytical ultracentrifugation (AUC) [59], dynamic light scattering (DLS) [60] and adsorption and intrusion techniques [5, 61], is crucial. Figure 1b exemplarily showcases how adsorption and intrusion techniques can be combined to assess both meso- and macropores in a porous silica particle sample.

In this study, we present a comprehensive workflow to investigate particulate materials by combined 360° ET, nanoCT and microCT, including a versatile preparation routine to create a conically shaped sample tailored to fit the different FOVs of each respective imaging technique (Figure 2). This allows us to investigate the size and morphology of several thousand particles as well as their internal pore system on the macro-scale, giving statistical relevance. At the same time, a tapered sample shape enables identical-location nanoCT investigations at higher resolution of smaller volumes containing only a few or even a single particle inside the same specimen. By acquiring data from matching ROIs through both techniques, a direct correlation between the data can be established, improving the potential and accuracy of quantitative analyses, which is further refined by utilizing neural network-based segmentation algorithms. As a model system, we take the example of porous silica particles prepared by polymerization-induced phase separation.[62-64] These particles exhibit a hierarchical pore structure that combines meso- and macropores within one particle.

## 2. Material and methods

### 2.1. Phase separated hierarchical porous particles

The porous particles are synthesized via polymerization induced phase separation.[62] In a typical synthesis, aqueous nitric acid (HNO3) solution is prepared by mixing 0.97 g HNO3 (65 %, VWR) with 10.26 g of de-ionized water generated by a water purification system (Purelab Flex 2, Elga Veolia). Then, 1.08 g of polyethylene glycol (PEG, 35000 gmol-1, Sigma-Aldrich) is dissolved in this solution. After complete dissolution, 8.32 g of tetraethyl orthosilicate (TEOS, ≥ 99%, VWR) is added with constant stirring (500 rpm). The solution is stirred until it becomes clear. The solution is then emulsified in perfluorinated oil (3M, Novec 7500) using a vortex shaker (2500 rpm, 20 s) to form droplets and kept for 3 days at T = 313.15 K for aging. The particles are filtered from the solution and washed with de-ionized water. They are then subjected to a solvent extraction treatment by immersion in

1M NH4OH solution for 9 h at T = 333 K. Afterwards, they are filtered again, washed with ethanol and allowed to dry at T = 333 K for 1 day. Finally, they are calcined at T = 1096 K for 6 h, with a heating rate of 1 Kmin-1. The target pore sizes for the resulting particles are desired to have macropores of ~5 µm and mesopores of ~15 nm.

*2.2. MicroCT and nanoCT sample preparation*

Ditscherlein et al. [65] developed a versatile preparation method of (nano-)particulate powders into columnar sample shapes optimized for microCT experiments. Firstly, the particles of interest are mixed with carbon black nanoparticles. Subsequently, the particle mixture is embedded in an epoxy resin matrix that is stable both mechanically and under vacuum and does not degrade when exposed to the X-ray beam. The carbon black nanoparticles act as spacer particles in the columnar samples, preventing the sedimentation of the particles of interest during the resin's curing process. Moreover, the carbon black particles together with the resin matrix exhibit low X-ray absorption and little change in refractive index, as shown in Table 1 for a X-ray energy of 5.4 keV as being used for the nanoCT experiments, to ensure sufficient contrast to the particles of interest in X-ray imaging. With the optimal mixing ratio, the spacer particles ensure that the particles of interest do not touch while maximizing their concentration in the columnar sample. The resulting sample allows for characterization of a large number of particles in one X-ray tomography measurement.

*Table 1: X-ray refractive indices n at 5.4 keV – the X-ray energy used for the nanoCT experiments – written as n = 1-δ-iβ, with δ as the real and β as the imaginary part of decrement of refraction [66], calculated from the real and imaginary part of the atomic scattering factor f [67], and transmission intensity through a thickness of 64 µm (field of view of the LFOV nanoCT imaging mode is 64 µm x 64 µm)[66] for materials relevant for this study.*

| Material | Refractive index at 5.4 keV  n = 1-δ-iβ | Transmission intensity | Attenuation length [µm] |
|---|---|---|---|
| **Carbon black (CB)** | $1 - 2.498 * 10^{-5} - i9.416 * 10^{-8}$ | 71.10 % | 187 |
| **Epoxy resin** | $1 - 8.388 * 10^{-6} - i9.170 * 10^{-8}$ | 72.16 % | 198 |
| **Epoxy & CB** | $1 - 8.712 * 10^{-6} - i9.176 * 10^{-8}$ | 72.14% | 196 |
| **Silica** | $1 - 1.657 * 10^{-5} - i4.461 * 10^{-7}$ | 19.88 % | 40 |
| **Oxygen** | $1 - 8.171 * 10^{-6} - i7.818 * 10^{-8}$ | 75.62 % | 229 |
| **Dry air** | $1 - 8.596 * 10^{-9} - i6.852 * 10^{-11}$ | 99.97 % | 263,000 |

The columnar samples used in this study are prepared according to the method outlined by Ditscherlein et al. [65] in the following manner:

For optimal separation, a mixture of 60 wt% (0.36 g) phase-separated silica particles (Figure 3a) to 40 wt% (0.24 g) carbon black spacer particles (Orion, Lamp Black 101, average particle size 95 nm) is found to be optimal. The components are gently mixed for 10 minutes on a sheet of paper using a rubber spatula and then transferred to a plastic container. Epoxy resin (Epo Thin 2, Bühler) is gradually added to the homogeneous mixture while gently stirring until it reached a honey-like consistency with a glossy surface (Figure 3b). Once the mixture achieves the desired consistency, it is slowly drawn into a silicon tube with an inner diameter of either 1 mm, 1.5 mm or 2 mm using a 1 ml syringe equipped with a shortened pipette tip. After filling, the cut silicon tube is straightened and affixed to a sheet of paper with sticky tape for curing. The columnar samples can then be extracted from the tube after 9 hours by cutting the silicon tube longitudinally with a razor blade and peeling it away using two fine-tipped tweezers. The final columns of different diameters are shown in Figure 3c.

To suit both the field of views of micro- and nanoCT measurements, these columnar samples have to be refined. The tips of the columnar samples are polished on four sides using silicon carbide paper (P1200 or P2400) to form a defined, pyramid-shaped tip, shown in Figure 3d. This geometry satisfies the size requirements of the different imaging modes described in Table 2. Investigations on the different thicknesses and the impact on the intensity transmission in nanoCT imaging can be found in Table S6 and Figure S15. In a last step, a single silica particle is glued (UHU BOOSTER, UV-glue) on top of the tip for contrast comparison between unfilled pores to silica and filled pores (resin) to silica (Figure 6, Figure S7). The particle is chosen via light microscopy and transferred on the glue coated tip via single-hair brush. The glue has similar X-ray properties as the epoxy resin (Table 2). Due to destruction of the original sample during transport after microCT analysis (Figure 6, Figure S3a-c), another, separate particle has been prepared on a steel needle for contrast comparison in LFOV nano-CT (Figure S7, Videos S5 and S12).

*2.3 360° ET sample preparation*

For 360° ET, a small pillar sample is prepared from one arbitrarily selected silica particle. To do so, the particle is first attached directly to a specialized 1 mm thin 360°-ET needle with tapered tip. For that, the needle's tip is coated with conductive silver adhesive (Acheson 1415) and then dipped into the particle powder. Using a stereo light microscope, the target particle's position is verified and adjusted as necessary to ensure its center position on the needle's tip. Once the adhesive had dried, the needle is transferred to an FEI Helios NanoLab 660 focused ion beam (FIB) and scanning electron microscope (SEM) for further thinning through ion milling. To align the region of interest and prevent structural damage from the macropores within the particle, the sample is milled at a 90° angle into a stepwise pyramid from opposite sides. A thin carbon coating is applied to the sample's top surface to reduce charging effects and additionally a charge neutralizer is employed during the milling process to minimize sample drift, which could lead to uneven milling. In the end, the top of the pyramid is subsequently milled into a pillar with a diameter of approximately 200 nm using stepwise cuts from the top.

*2.4. MicroCT*

The microCT datasets are acquired with a Zeiss Xradia 620 Versa X-ray microscope. It is based on a classical cone beam computed tomography setup in combination with an optical lens system to further increase the achievable resolution. The X-rays are generated by a tunable polychromatic source with tungsten as target material, where the acceleration voltage can be varied between 30 kV and 160 kV at a power up to 25 W. This enables a wide range of materials to be analyzed, i.e. starting from polymers up to high Z materials like metals.[68] On the detector side, the X-ray photons are converted to visible light by a scintillator. The resulting image is magnified by the optical microscope and recorded by a sCMOS camera consisting of a 2048 x 2048 pixel matrix with a physical pixel size of 11 µm each. Depending on the sample and the desired resolution, there are in total four optical magnifications available (0.4x, 4x, 20x, 40x). Furthermore, source and detector are mounted on one linear axis each to adjust the factor of geometric magnification. Through this setup, the FOV can be varied, based on the geometric and optical magnifications, achieving a spatial resolution of up to 0.5 µm [68]. The sample is mounted on a special sample holder designed by Zeiss and is placed on a stage with one rotational and three linear axes. The scan parameters, like exposure time, voltage, power and filter are set depending on sample material, transmission and signal-to-noise ratio (SNR). For rotationally symmetric samples, the projections for the CT are acquired during a 360° rotation of the sample, where the number of projections is uneven to avoid redundancy. Afterwards, the reconstruction is performed automatically using a Feldkamp-David-Kress (FDK) algorithm and applying an adaptive motion compensation to correct source drift, thermal drift and sample shifts.[68] The scans with 250 nm and 1000 nm effective pixel size are acquired with a source setting of 40 kV and 3 W, while the 500 nm scan uses a setting of 50 kV and 2.5 W. The effective pixel size results from the geometric and optical magnification and represents the size of a pixel in the resulting projection. It can be calculated by dividing the physical pixel size of the camera by the product of optical and geometrical magnification. As the sample is low absorbing in the used energy range, no source filtering is required. Moreover, the proprietary mode for the propagation-based phase contrast of the Zeiss Xradia 620 Versa has been utilized. The exposure time is increased with the sampling becoming finer to keep the intensity on a constant level. This is necessary, as for each of the three pixel sizes, a different objective had to be used and the intensity on the camera decreases by increasing the optical magnification. The in-depth imaging parameters, such as resolution and binning, for all microCT measurements can be found in Table 2. The nominal resolution is the resolution given by the manufacturer (Zeiss), whereas the estimated resolution (Table S5) is an in-plane resolution determined via the derivate of intensity line profiles, as described in the Supporting Information and visualized in Figure S12. The reconstructions are performed using the Zeiss reconstruction software, while the values for horizontal shift and beam-hardening-correction are set manually. The sub reconstructions required for the Fourier shell correlation are done with the same reconstruction settings in which only even or odd numbered projections are used.

*2.5. NanoCT*

For nanoCT characterization, a Zeiss Xradia 810 Ultra X-ray microscope (XRM) equipped with a rotating chromium anode is utilized, supplying in the imaging setup X-rays of quasi-monochromatic 5.4 keV. It offers two different contrasts modes: (1) absorption contrast, which is based on X-ray absorption described by Beer-Lambert law, and (2) Zernike phase contrast. Here, a phase ring is introduced into the beam path behind the Fresnel zone plate. Further, two resolution modes are available. The large-field-of-view (LFOV) mode offers a field of view of 64 µm x 64 µm with a nominal spatial resolution of down to 150 nm and a pixel size of 65 nm on the 1024 x 1024 pixel CCD sensor.[29] On the other hand, the high resolution (HRES) mode offers a nominal spatial resolution of down to 50 nm with a field of view of 16 µm x 16 µm and a pixel size of 16 nm.[29] A columnar sample is first mounted onto the "SEM Stub" sample holder for the Zeiss Xradia Ultra products and then placed onto the stage of the XRM. Then, the ROI for the scan is chosen, depending on the imaging mode. Exposure time is determined in relation to contrast and SNR, depending on the sample and contrast mode. Further, the tilt-angle range is set to full 180° and binning to 1. Forty reference images without sample are acquired before and after tilt series acquisition, respectively. These reference images are averaged and used for a bright-field gain correction of the detector as well as to flatten the profile of the X-ray beam. This is done by normalizing the averaged reference image to 1 and dividing each projection by this normalized reference. The drift correction is set to adaptive motion compensation (AMC), an image alignment post-processing routine implemented in the Zeiss "Scout&Scan" software (version 13.0.6602.39366). The number of projections is selected via the Crowther Criterion to match the desired resolution.[68] For samples exceeding the FOV (so-called "interior tomography") [69], a higher projection number can be chosen to suppress reconstruction artefacts [70]. The impact of interior tomography on nanoCT contrast can be seen in Table S6 and Figure S15. The imaging parameters for all measurements can be found in Table 2.

*Table 2: Imaging parameters for all tomography measurements described in this study.*

| Imaging mode | Microscope | # of Projections | Exposure Time per Frame [s] | Binning | Tilt Range | Resolution [nm] Nominal | Resolution [nm] Estimated | Resolution [nm] FSC ½-bit | Pixel Size [nm] | Projection size [px x px] | FOV Size [µm x µm] |
|---|---|---|---|---|---|---|---|---|---|---|---|
| Low resolution µCT (4x) | Xradia Versa 620 | 1601 | 21 | 1 | 360° | 1900 | 2520 | 2207 | 1000 | 2048x2048 | 2039x2039 |
| Medium resolution µCT (20x) | Xradia Versa 620 | 1601 | 24 | 2 | 360° | 900 | 1520 | 1031 | 500 | 1024x1024 | 504x504 |
| High resolution µCT (40x) | Xradia Versa 620 | 1601 | 48 | 2 | 360° | 500 | 928 | 774 | 250 | 1024x1024 | 252x252 |
| LFOV nanoCT | Xradia Ultra 810 | 1601 | 80 | 1 | 180° | 150 | / | 208 | 65 | 1024x1024 | 64x64 |
| HRES nanoCT | Xradia Ultra 810 | 1801 | 120 | 1 | 180° | 50 | / | 138 | 16 | 1024x1024 | 16x16 |
| Electron tomography | FEI Titan³ Themis | 180 | 10.43 | 1 | 180° | 0.4 | / | 2.2 | 0.54 | 2048x2048 | 1.1x1.1 |

The nanoCT data is reconstructed via a simultaneous iterative reconstruction technique (SIRT) algorithm[71] implementation based on the Astra Toolbox[72-74]. This implementation, as described in more detail by Englisch et al.[29], is applied using a reconstruction routine containing a cosmic radiation correction, reference correction, AMC-based drift correction and center-shift correction. The number of SIRT iterations is set to 100 iterations for all datasets.

*2.6 360° ET characterization*

A 360° electron tomography analysis in high-angle annular dark-field (HAADF) scanning transmission electron microcopy (STEM) imaging mode is performed on the pillar sample with a double Cs-corrected FEI Titan³ Themis 60-300 operated at 300 kV, equipped with a Fischione Model 2050 On-Axis Rotation Tomography holder. To reduce potential beam damage to the silica pillar, images are collected with a beam current of 30 pA, a dwell time of 5 µs per pixel, and a pixel size of 0.54 nm, resulting in the dose of 32 e$^-$/Å$^2$ per single tilt. The data is recorded with 2048 x 2048 pixels and a tilt increment of 1°. Auto-focusing is performed on a region adjacent to the imaging area to preserve sample integrity. A convergence semi-angle of 3 mrad is employed using a custom 10 µm C2 aperture in aberration-corrected nanoprobe mode. Further imaging parameters can be found in Table 2. A stack alignment of the tilt series is performed, based on the center-of-mass projections algorithm[75] implemented in the "ToReAl" Matlab script. Reconstruction of the 360° ET data is completed using a SIRT algorithm with 50 iterations, as detailed in prior works[72-74].

*2.7. 3D data analysis*

An overview of the different 3D analysis methods can be found in Table S2. Any resulting data is plotted in Origin (OriginLab, version 2022 9.9.9.225) and fit curves of a suitable function applied for any histogram are generated via the "nonlinear curve fit tool" and fit curves with multiple peaks via the "Multiple Peak Fit Tool".[76] The tilt series shown in Figure 4 are visualized via ZEISS XMController, a proprietary software of the Zeiss Xradia 810 Ultra.

Visualization and registration of the 3D reconstruction data (Figure 5) are performed using arivis Vision4D version 4.1.1. The software allows for convenient post-processing and visualization of datasets with live previews of post-processing results, simplifying the registration of different datasets. For the latter, a slice of the LFOV nanoCT reconstruction is selected as a reference and the other reconstructed datasets are rotated until the same sample orientation is achieved, allowing a direct correlation of the datasets.

*2.7.1. Spatial resolution and interface sharpness determination*

The resolution of the 3D reconstructions was investigated via Fourier shell correlation (FSC). Here, two sub-reconstructions are created by using the separated even and odd numbered projections from the original tilt series, Fourier transformed and correlated. From the resulting correlation function, a value for the spatial resolution can be determined via a threshold curve (see Figure S4). For a more conservative value, the 1-bit criterion curve can be applied but the ½-bit criterion has proven to be the more accurate.[29, 77]

Further, the sharpness of the interface between silica and resin are determined for both microCT and nanoCT reconstructions, and compared to the interface silica/air in nanoCT via proximity histograms or, in short, proxigrams. Conventionally, proxigrams are used in atom probe tomography applications to determine a profile of local atomic concentration against distance (proximity) to an interface.[78, 79] In this work, the change of intensities at phase boundaries (i.e. pore to silica) is investigated instead, shedding light onto segmentation and reconstruction quality. A customized version of the code from Felfer et al. [80] was implemented utilizing Matlab (MathWorks, version R2023a) to generate the proxigrams. To do so, the segmentation of a dataset is used to determine the position of the interface in the reconstruction, which is then used to generate an isosurface (Figure S5a,b). Subsequently, the averaged intensity profile is calculated perpendicular to the isosurface in the selected volume (Figure S5c,d). The resulting intensity profiles are plotted as the normalized intensity (normalized to maximum extracted intensity as 1 and minimum extracted intensity as 0) over the distance to the isosurface in microns (Figure S5d). The proxigrams are evaluated regarding the positions at 25% ($x_{25\%}$) and 75% ($x_{75\%}$) of normalized intensity as well as regarding the slope in between these x-positions via a Python code (Python software foundation, version 3.12.7) implemented into Spyder (Spyder IDE, version 5.5.1). The slopes of the intensity profiles are investigated as a measure of the sharpness of the interfaces, whereas the distance (Δx) between $x_{25\%}$ and $x_{75\%}$ can be extracted as a lower resolution limit (Table 3) [81]. Finally, the difference Δy between the local maxima and minima of the normalized intensity plateaus left and right of the interface is calculated (Table 3, absolute intensity difference values in Table S4) as a measure for the intensity of phase contrast artefacts (Figure S6a and Figure S7), after extracting the corresponding y-values via Origin (OriginLab, version 2022 9.9.9.225).

*2.7.2. Segmentation of LFOV nanoCT and ET datasets*

For quantitative analysis, the LFOV nanoCT reconstruction (Figure 5e) is segmented into particle and pore (i.e., epoxy matrix) space by utilizing arivis' marker-based machine learning implementation of a random forest algorithm based on Ilastik [82, 83]. This creates a stack of 16-bit grayscale intensity images, after which the image is binarized where background (i.e., epoxy resin including carbon black particles) is mapped to black pixels (grayscale value 0) and particle solid is mapped to white pixels (uniform grayscale value of >0). Before segmentation, a spherical mask is applied to the particle, removing the rough surface and the background surrounding the particle. This leads to an increase in segmentation and pore analysis quality as background and rough surfaces are prone to introducing errors, inaccuracies and additional artefacts. For further processing, these image stacks are transformed into 8-bit data, maintaining the grayscale value of the black pixels (0) while setting the white pixels grayscale values to maximum 8-bit value (255).

The reconstructed ET data is first transformed into 8-bit, then, a Gaussian blur with a radius of 5 pixels is applied before segmentation to reduce noise and ensure smooth pore walls. Manual thresholding is then applied to a value chosen manually to segment the data (here 70), based on visual quality of the segmentation, leading to the same type of segmentation dataset as from LFOV nanoCT.

*2.7.3 Quantitative pore analysis of LFOV nanoCT and ET datasets*

The pore sizes of the particulate sample are analyzed in the segmented LFOV nanoCT and ET reconstructions via the "local thickness" plugin of the open-source platform ImageJ.[84, 85] The plugin performs a maximum sphere inscription (MSI) by assigning every voxel a value relating to the size of the pore (short axis of the pore) containing the voxel.[84-87] To transform this into a discrete pore size distribution, the relative number of voxels (i.e., dividing the number of voxels assigned to a certain pore diameter by the total number of pore voxels) is plotted over their assigned pore size. The peak pore diameter (mode) is then extracted via a non-linear log-normal fit over the distribution.[5] Further, the mean pore size and the standard deviation are extracted from the distribution. The according standard deviation is written as error to the mean values.

To determine the macropore porosity in the LFOV nanoCT reconstruction, a total pore volume is extracted from the corresponding pore size distribution. The MSI process is repeated for the particle solid, enabling the extraction of the solid volume. By combining the pore and solid, a total sample volume is extracted. The porosity is now calculated by dividing the total pore volume by the total sample volume. To calculate the corrected value for the porosity including both meso- and macropores (see Chapter 3.6), the solid volumes determined from the LFOV nanoCT and microCT reconstructions are reduced by the mesopore volume determined by the ET analysis, which is further added to the total pore volumes.

*2.7.4 Segmentation and quantitative pore and particle analysis of microCT dataset*

For analysis of the lowest-resolution microCT dataset, the software XamFlow (Lucid Concepts, version 1.10.2.1.) was utilized on which machine learning and classical image processing algorithms can be applied to create an annotation-efficient analysis pipeline in three consecutive workflow steps:
(1) Preparation of 2D-training data and U-Net training
(2) Preparation of 3D-training data and U-Net training
(3) Final segmentation and quantitative analysis

(1) First, the reconstruction is imported to be available for the analysis workflow shown in Figure S8. Then, semi-automatic labelling in 2D is performed. Here, ten sub datasets of 200 x 200 x 200 voxels in size are extracted and a 3D Gaussian filter applied, with a kernel size of 5 x 5 x 5 and a sigma of 1.2. From each of these sub datasets, up to five slices are cropped down further and labelled, utilizing hierarchical labelling. Then, the initial segmentation is performed on each of the slices by utilizing thresholding and filling any cavity inside the segmented particles and any residing errors are corrected with a manual brush tool. This results in masks covering the particles including the pore space, which are utilized to extract outer shells of the particles via a distance transform. Further, the masks are utilized to remove the background, retaining only the pore space and solid of the particles. Via additional thresholding, the solid is extracted. Finally, by inverting the masked segmentation, the pore space is extracted. Due to the interface resin-air exhibiting phase contrast artefacts, air bubbles have to be removed or separated from the solid and pore space of the particles. Therefore, an additional thresholding step is performed, followed by a dilation modifier to ensure the full bubble is included. The studied material is a multi-material composite consisting of extracted phases, such as silica, air, intraparticle pores, and background. These phases can be distinctly segmented and labeled in the images, with air bubbles assigned as regular labels, while particles, particle shells, and pore spaces are categorized using hierarchical labels. In the final step of (1), the pairing of Gauss filtered slices and labels is used as input for a 2D nn-UNet model[88], using a single fold and 100 epochs for training (Figure S9a).

(2) For the second workflow step (Figure S10), similar to (1), sub-datasets are taken from the reconstruction. Here, 19 datasets à 256 x 256 x 256 voxels are generated and filtered with the same Gaussian filter as in (1). Now, the trained 2D nn-Unet is used for slice-by-slice segmentation and resulting segments are filtered with a 3D median filter of a 3 x 3 x 3 kernel to remove 2D artefacts. Subsequently, a morphological Opening modifier[89-91] of a 3-pixel distance is applied to disconnect any touching particles. These segments are utilized as particle mask after visually verifying and correcting them with the manual 3D brush tool. From these masks, a two-pixel shell is generated by utilizing a one-pixel erosion followed by a 3D distance transformation. Subsequently, masks for the bubbles are generated using a pipeline for robust segmentation that is less complex, compared to the utilized nn-Unet for the particles. Now, the generated masks are applied again as labels with the air bubbles as regular labels and the particle and particle shells as hierarchical labels. However, pore space is not added as a separate label as generation of the corresponding masks is quite straight forward and simple, again utilizing thresholding and inversion on the particle masks. Finally, a 3D nn-UNet model[88] is trained with a single fold and 100 epochs (Figure S9b), utilizing the pairing of the 3D sub-datasets and their 3D hierarchical labels.

(3) In the final workflow step (Figure S11), the full image is segmented with the trained 3D nn-Unet model and the results filtered utilizing thresholding of a strongly Gaussian filtered image to remove the low amount of false positive detections. The resulting particle masks, covering the whole particle volume, are revised, filling inner cavities by utilizing the component labeling of the background. Subsequently, fused particle masks are disconnected via a 3D morphological Open modifier of a 2-pixel distance. For determining a measure for the particle diameter, a spherical shape is assumed and the diameter of the volume equivalent sphere, or equivalent spherical diameter (ESD)[92], is calculated after extracting the volume of each 3D particle mask via 3D connected component labeling. The resulting diameters are mapped as a color scale on each particle in a 3D visualization. The pore space is extracted by masking a background mask, containing everything but the solid of the particles, with the particle mask. From this and the particle mask, the solid of the particles can be extracted. Finally, the porosity per particle is calculated via a component labeling based method and the pore size distribution extracted via a local thickness implementation.[86] A histogram for pore sizes, particle sizes and per particle porosity is created. The according standard deviation is calculated from the histogram data and written as error to the mean values.

## 3. Results and discussion

### 3.1. Correlative workflow

The correlative workflow in Figure 2 starts with embedding the particles together with carbon black spacer particles in a resin column as described in Chapter 2.2. This allows for 3D characterization of large quantities of well-separated particles facilitating segmentation. The second step is creating a conically-shaped tip by, e.g., mechanical grinding and milling or by laser ablation. This sample geometry allows for subsequent flexible microCT and nanoCT analysis covering multiple FOVs and resolutions. While microCT enables 3D analysis of a large amount of particles allowing for good statistics, nanoCT is applied for high-resolution 3D scans of smaller features. The conical geometry allows for imaging congruent ROIs in the differently magnified tomographic acquisitions, which is crucial for a direct volume correlation and scale-bridging feature analysis.

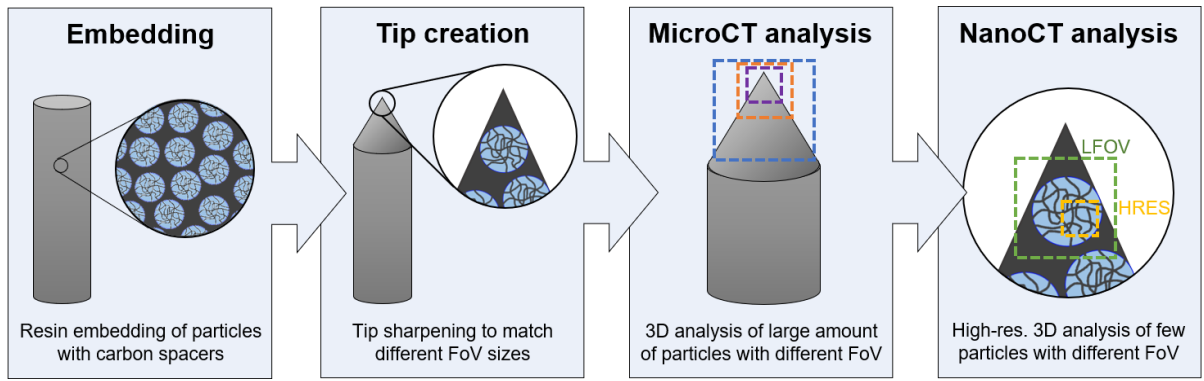

*Figure 2: Scheme of the correlative and scale-bridging workflow. Starting with the embedding of particles into a resin column, allowing for microCT characterization of large quantities. Carbon spacer particles are added to the resin matrix to simplify separation and segmentation of the silica particles. In step two, a conically-shaped tip is created adapted to the different field of view (FOV) sizes of nanoCT and microCT. Then, the sample is characterized using microCT at different magnifications. Here, it is of high importance to fully image the sample including the tip end, since this region is used for direct nanoCT correlation in the final step. For the latter, a single silica particle is imaged in LFOV nanoCT mode and a magnified region of interest (ROI) of it in HRES nanoCT mode.*

*3.2. Sample preparation*

To ensure a sample being suited for both micro- and nanoCT measurements, its geometry and dimensions must match the varying FOVs of the different X-ray microscope setups (see Figure 2 and Table 2 for an overview of the different FOVs used in this study). This alignment is essential for achieving optimal imaging conditions and avoiding interior tomography artifacts.[69, 93, 94] Moreover, the X-ray attenuation of the specimen needs to be considered, with a rule of thumb being that the sample thickness may not exceed twice the X-ray's absorption length (Table 1).[15] To satisfy the requirements for the different modalities utilized in this study, the tip of a columnar sample, measuring 1.5 mm in diameter, is polished on four sides using silicon carbide paper to form a well-defined, pyramid-shaped tip, as visible in Figure 3d. This design enables the upper portion of the tip to be used for identifying suitable individual particles for nanoCT measurements, while the entire columnar sample, including the tip, can be imaged using microCT.

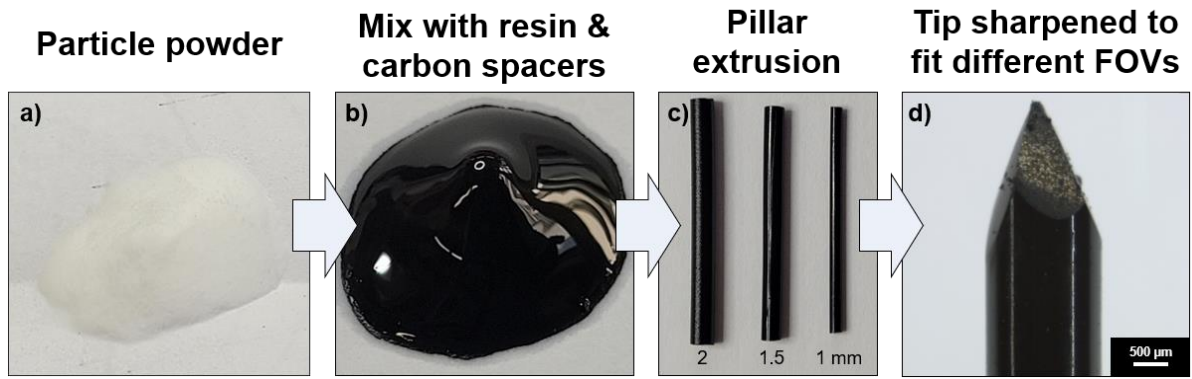

*Figure 3: Steps in sample preparation: (a) Particles of interest in powder form (here phase-separated silica particles). (b) Mixture of silica particles with carbon black spacer particles and epoxy resin. (c) Resulting columnar pillar samples of different diameters. (d) Final sharpened sample, fitting the different field of views of the applied imaging modalities. Particles on the surface (bright contrast) can already be seen under the light microscope.*

*3.3. MicroCT acquisition*

After preparation, the sample is first analyzed using microCT at three different resolution settings (Table 2). Beginning with a low-resolution scan at a pixel size of 1000 nm, the complete conical shaped sample was scanned, see Figure 4a and Video S1. The scan shows thousands of embedded particles as dark spots surrounded by the brighter contrast corresponding to the resin matrix. This pattern is regularly interrupted by bright spots, resulting from air bubbles trapped inside the resin. The tip itself is surrounded by glue, stabilizing a single, unfilled particle on top of the tip, which was planned to be used for LFOV nanoCT contrast comparison between unfilled pores to silica and filled pores (resin) to silica (Figure 6, Figure S3c). The consecutive intermediate- (Figure 4b, Video S2) and high-resolution (Figure S3c, Video S3) scans with corresponding 500 nm and 250 nm pixel size were performed focusing more and more on the tip of the sample, where subsequently nanoCT characterization is performed (cf. Figure 2). The intermediate scan captures hundreds of particles in projection with fewer air bubbles and provides decent resolution (Figure 4b, Video S2), while the high-resolution scan focuses on only a few dozen particles, which are clearly resolved (Figure S3c, Video S3). Figure S3 gives a complete overview over all acquired tilt series.

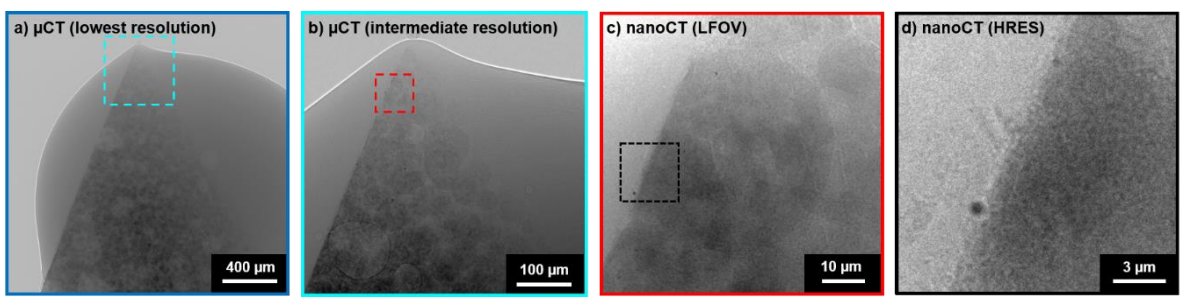

*Figure 4: Exemplary projection images from the microCT and nanoCT tilt series (see Figure S3 for a complete overview): a) Lowest and b) intermediate resolution microCT tilt series; c) LFOV nanoCT tilt series of a single particle; d) HRES nanoCT tilt series from a sub-region of the particle from c). Detailed imaging parameters for these tilt series can be found in Table 2 and Videos S1 – S6 show animations of the full tilt series.*

*3.4. NanoCT acquisition*

After microCT characterization, the sample is investigated at higher resolution using nanoCT. Starting with the LFOV nanoCT mode, a suitable particle is identified (Figure 4c) in the thinner part of the larger sample volume by acquiring mosaic XRM projection scans (Figure S1) in front and side view covering an in total larger FOV. After confirming that the particle and tip thickness are suitable, a LFOV nanoCT tilt series is acquired (Figure 4c, Video S4) with parameters detailed in Table 2. Here, the selected particle of interest is visible, but, due to the sample being larger than the FOV at selected ROI, objects outside of the ROI are visible in projection as well. Moreover, due to the tapered geometry, the thickness of the sample rises significantly along the vertical axis from top to bottom) of the FOV, leading to an intensity gradient in the image. This gradient can also be seen in the corresponding reconstruction shown in Figure 5e and Video S11. The reconstructed dataset is investigated to find interesting ROIs for subsequent HRES nanoCT analysis. This was done by virtually scanning different ROIs of the LFOV nanoCT dataset by cropping it down to cubes with 16 µm side length, corresponding to the covered volume in HRES mode. In the studied sample, one fissure and several smaller pores were identified and chosen as ROI for HRES investigation (Figure S2), as indicated in Figure 5e. An important requirement for HRES nanoCT is the sample thickness: if the ROI for HRES nanoCT is not directly located at the very tip of the pillar, the sample size might exceed the FOV of the imaging mode. This situation is usually referred to as interior tomography, which, on the one hand, leads to specific 3D reconstruction artifacts, and, on the other hand, may worsen the SNR in the projection image (Figure 4d) and also the corresponding 3D reconstruction (Figure 5f). An analysis of the dependence of measured intensity on the local sample size exceeding the FOV in nanoCT imaging is provided in Table S6 and Figure S15. In this analysis, the local sample composition along multiple lines in the lowest-resolution microCT reconstruction was estimated, and the corresponding transmission intensities were calculated. The interior tomography conditions hamper proper alignment of the XRM stage's rotation center (eucentric position) towards smaller features of interest, which might be difficult to identify in the HRES nanoCT projections (Figure 4d). Instead, the HRES nanoCT ROI needs to be localized relative to distinctive, larger features of the sample, such as features inside or on the surface of the selected particle, the surface of the tip, or surrounding particles. Here, a group of pores (Figure S2) was chosen to identify the correct FOV corresponding to the selected ROI. To validate the correct positioning of the XRM stage and its rotation axis, a preliminary tilt series with less projections and exposure time was acquired. This step allows for subsequent repositioning of the sample, if necessary, to ensure 3D characterization of the correct ROI, being located in eucentric position of the HRES mode. This procedure enables the ROI localization within the LFOV nanoCT reconstruction with a precision of approximately 1 µm. After repositioning of the sample to the selected ROI, a HRES tilt series was acquired (Figure 4d, Video S3) with the parameters described in Table 2. Due to interior tomography imaging conditions of a volume with smaller dimensions than the thickness of the sample, the SNR of this tilt series is reduced. The corresponding 3D reconstruction is shown in Figure 5f and Video S13. Here, even though the SNR is also rather low, the desired features (cracks and smaller pores) are still well resolved.

*3.5. 3D data analysis*

Virtual slices from the 3D reconstructions of the five acquired datasets (Table 2) are depicted in Figure 5 with the correlating positions of higher resolution scans indicated by colored rectangles. In comparison to the tilt series in Figure 4, the contrast is now inverted, a result of the image processing during reconstruction. Now, components get brighter with rising density or thickness. Figure 5a shows a slice through an overview microCT reconstruction (largest FOV of 2038 µm x 2038 µm; see Table 2) of the sample tip with the whole reconstructed volume containing 14,553 individual particles. Further, the air bubbles trapped inside the resin are visible as dark spots (vice versa bright spots in Figure 4a). This allows analyzing a statistically relevant number of particles of the embedded particle batch with respect to, e.g., its particle size distribution (Figure 7a), providing important feedback to synthesis. Moreover, larger macropore fractions can be investigated (Figure 7b), albeit at limited resolution (about 2.2 µm, see Figure S4a and Table 2). Two additional microCT scans of the same specimen from the ROIs indicated in Figure 5a were acquired at intermediate and high resolution. Figure 5b shows a slice through the intermediate microCT reconstruction with a FOV of 504 µm x 504 µm, offering insights with higher fidelity into the particulate structure. Smaller particles, particle fragments and also smaller intraparticle pores become apparent. This scan allows for a more detailed characterization of the particles while maintaining a reasonably good statistical relevance, containing 613 particles. The microCT scan with highest resolution and smallest FOV of 252 µm x 252 µm is shown in Figure 5c. Only a relatively small amount of 53 particles is contained in this volume. However, previously unseen details become visible, e.g., larger agglomerates of carbon black particles inside the epoxy matrix between the silica particles can partly be discerned. More importantly, as visible in Figure 5d, when zoomed into individual particles, the interior pore space of the silica particles is resolved in detail. This high-resolution microCT mode poses as a well-suited link between micro- and nanoCT, exhibiting an about 4x larger FOV and 6x lower resolution compared to a LFOV nanoCT scan, enabling precise registration and correlation of the different datasets. The LFOV nanoCT mode with a FOV of 65 µm x 65 µm offers an even higher resolution, resolving smaller features down to about 150 nm such as cracks or smaller pores (Figure 5e). This allows for an in-depth characterization of the pore system and particle features of one complete, individually selected silica particle. Figure 5f shows a slice through the HRES nanoCT reconstruction with a FOV of 16 µm x 16 µm. This slice displays the area indicated in the LFOV nanoCT reconstruction in Figure 5e. Due to the higher spatial resolution of about 50 nm, finer structures such as smaller pores or fissures are observed, exclusively being resolved using this 3D imaging mode. When correlating and comparing the nanoCT and microCT reconstructions (Figure 5 and Videos 8-13), the representative spherical shape and uniform pore system of the selected particle for nanoCT investigations become apparent.

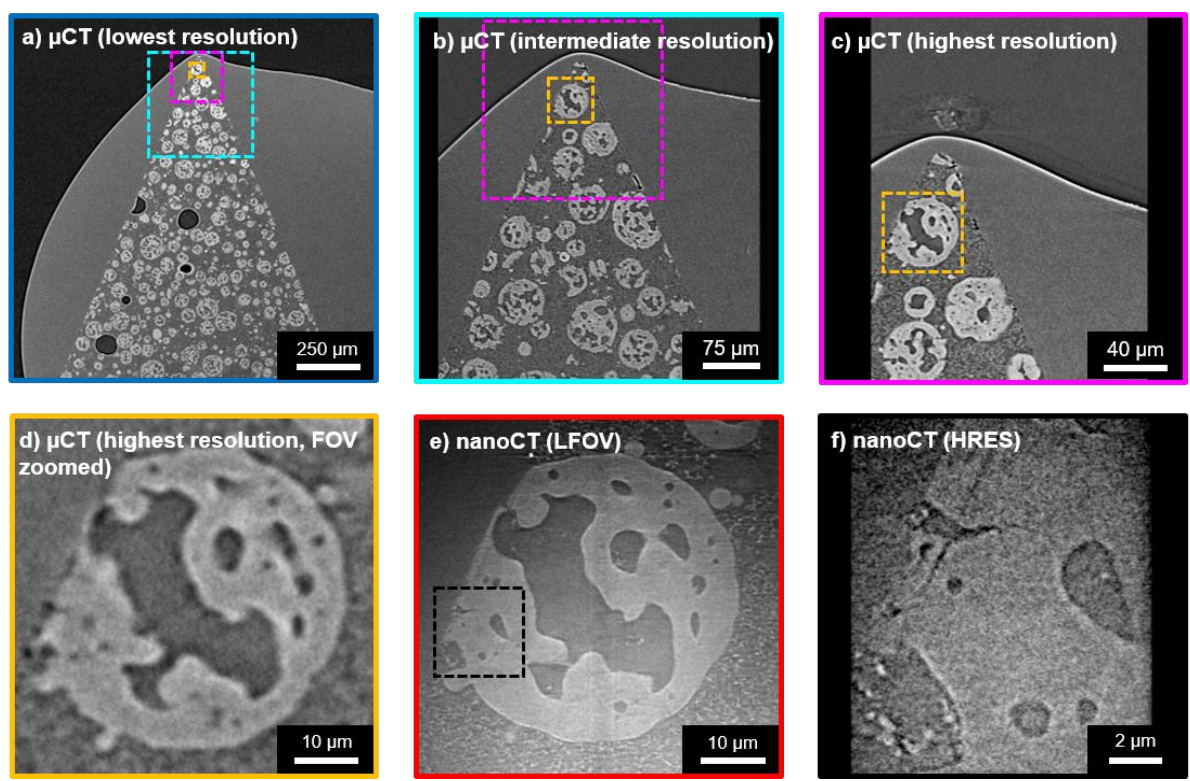

*Figure 5: Correlative tomography of the particulate sample: (a) Lowest-resolution microCT reconstruction of a large volume containing 14,553 silica particles. The FOVs of (b), (c) and (d) are marked with teal, magenta and orange boxes. (b) Intermediate-resolution microCT reconstruction. The FOVs of (c) and (d) are marked in magenta and orange. (c) High-resolution microCT reconstruction, the FOV of (d) is marked in orange. (d) Zoomed-in excerpt of a selected particle in (c) highlighting the same ROI as imaged in (e) using large-field-of-view (LFOV) nanoCT with (f) high-resolution (HRES) nanoCT marked in black.*

To investigate the quality of the nanoCT and microCT reconstructions, the spatial resolution in the reconstructed volumes was quantified by Fourier shell correlation (FSC) (see Figure S4a-e, Table S3 and Table 2). The extracted resolutions for the ½-bit criterion are 138 nm for HRES and 208 nm for LFOV nanoCT, whereas the microCT resolutions are 774 nm for 250 nm pixel size (highest resolution), 1031 nm for 500 nm pixel size (intermediate resolution) and 2.2 µm for 1 µm pixel size (lowest resolution), respectively. For the microCT volumes, the calculated FSC resolution is inferior to the nominal resolution

(Table 2), as the scan parameters for microCT were not chosen to achieve the highest possible spatial resolution, but rather to cover a significant FOV for each imaging mode. The reduced resolution in the nanoCT reconstructions, both for HRES and LFOV, can be explained by artefacts and low SNR due to interior tomography conditions, thus the influence of the surrounding sample features outside the ROI (cf. Figure 5e,f and Videos S11 and S13).

To further examine the quality and resolution of the nanoCT and microCT reconstructions, we determined the interface sharpness between pore space and silica material by calculating proxigrams from a selected volume of (300 px)³ in the center of the datasets (see Figure S5). The proxigrams in Figure 6 show a clear intensity gradient when transitioning from the pore space (resin or air) to the silica structure. As expected for a higher resolution reconstruction, the LFOV nanoCT volume exhibits a steeper gradient with a slope of 3.35 $^1/_{\mu m}$ in the silica/resin interface compared to even the highest-resolution microCT dataset, with a slope of 0.91 $^1/_{\mu m}$, emphasizing that nanoCT resolves the interface sharper. The intermediate and lowest-resolution microCT data fall in line with that, showing a slope of 0.67 $^1/_{\mu m}$ and 0.39 $^1/_{\mu m}$, respectively.

To investigate contrast differences between the silica/air and silica/resin interface, an unfilled particle was glued on top of the sharpened tip with the resin-embedded particles, and reconstructed in 3D together with the embedded particles (Figure S3a-c). As the particle was destroyed during transport prior to LFOV nanoCT investigation, a separate particle was prepared on a steel needle, imaged and reconstructed with the same parameters and used for contrast comparison instead, see Figure S7 (Videos S5 and S12). In the LFOV nanoCT reconstruction, the silica/air interface exhibits a slope of 4.32 $^1/_{\mu m}$ and is therefore steeper than the slope of silica/resin. The reason for this lies in the different refractive indices (Table 1) and the corresponding edge-enhancing phase contrast. Due to the higher difference in refractive indices of silica/air than silica/resin, the phase contrast enhancement located at the interface is more pronounced. Further, the interface resolutions of the datasets were evaluated by extracting $x_{25\%}$ and $x_{75\%}$ values to calculate their difference values $\Delta x$ (Table 3, see Figure S6a for an explanation), extracted from the graphs of Figure 6 (see also Figure S6), where these values are indicated with orange, horizontal dotted lines at 25 % and 75 % of normalized intensity. For the microCT silica/resin interface, we determined an interface resolution of 1.31 µm for the lowest, 0.76 µm for the intermediate and 0.46 µm for the highest magnification, so values below both nominal and FSC resolution (Table 2), indicating that the interface silica/resin is resolved better than the FSC indicates. For the LFOV nanoCT silica/resin interface, an interface resolution of 0.14 µm was determined, undercutting the resolution by FSC and lying closely below the nominal resolution (Table 2). This implies that the interior tomography conditions influence image quality, however generating a high-quality segmentation of the interface silica/resin in the LFOV nanoCT dataset is still possible. Finally, for the LFOV nanoCT silica/air interface, a resolution of 0.12 µm was determined, a value below the nominal and FSC resolution, and also slightly below the interface resolution of the LFOV nanoCT silica/resin dataset. This indicates that both the silica/air and silica/resin interfaces are well resolved, although the silica/air interface demonstrates slightly better resolution. This is primarily due to the stronger phase contrast and edge enhancement at the silica/air boundary. Additionally, the non-embedded particle is not surrounded by other parts of the sample or adhesive, preventing interior tomography artifacts from affecting the reconstruction quality.

However, when comparing the intensity profiles in Figure 6 (right), the silica/air profile (brown curve) shows a decrease in intensity after the maximum intensity (solid phase) and an increase in intensity before the minimum intensity (pore space) when compared to the silica/resin profile (red curve). This indicates that the intensity difference between the bulk phases—pore space (left) and solid phase (right)—is lower for silica/air than for silica/resin. While the interface is steeper for silica/air, the overall contrast between the bulk phases is reduced. This can be attributed to the higher refractive index difference between silica and air compared to silica and resin. For the silica/air interface, edges appear more pronounced evolving in so-called "halo" artifacts, while the contrast between bulk and pore regions is more uniform, resulting from "shade-off" artifacts.[95-97]

To quantify the contrast difference in the bulk phases, the $\Delta y$ values (see Figure S6a for an explanation) between the minimum and maximum intensities of the normalized intensity plateaus in the bulk were extracted (Table 3, for absolute intensity difference values refer to Table S4). In the microCT profiles shown in Figure 6 (left), a decrease in $\Delta y$ is observed when moving from the lowest resolution, with a normalized intensity difference of 0.661, to the highest resolution, with a difference of 0.475. This indicates a higher influence of phase contrast artifacts at higher resolutions, as a lower $\Delta y$ corresponds to reduced contrast between pores and silica and a higher presence of shade-off artifacts. For the LFOV nanoCT profiles, the silica/air boundary shows a $\Delta y$ of 0.037, while the silica/resin interface exhibits a $\Delta y$ of 0.454. This more than tenfold increase in intensity difference between the pore space and silica after embedding indicates a significant reduction in phase contrast artifacts and allows for a much better distinction between pores and solid, enhancing segmentation quality and robustness. This contrast increase outperforms the slightly steeper silica/air interface, so that the particle embedding does not impair imaging conditions but rather improves the image quality.

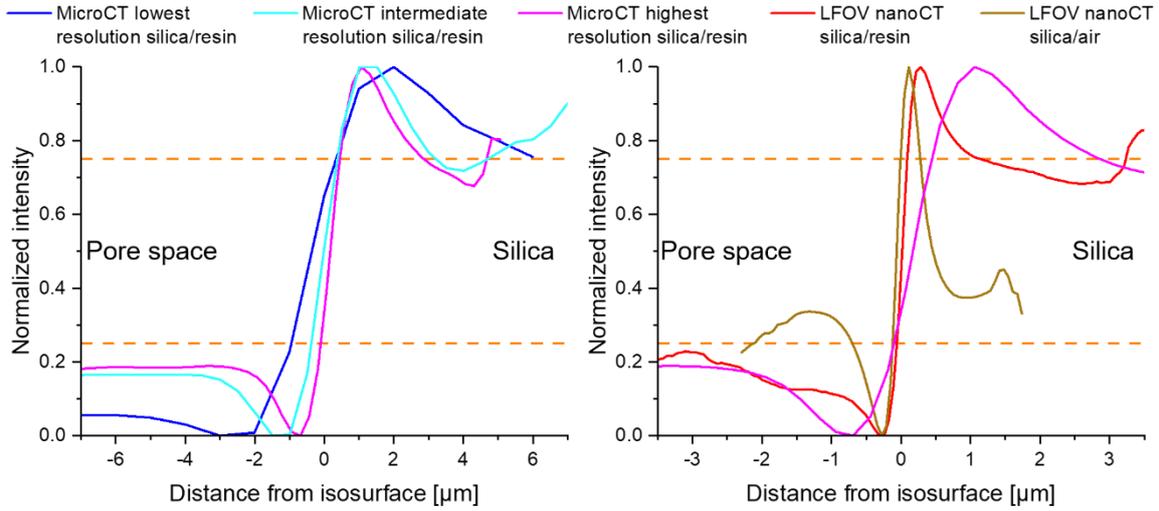

*Figure 6: Comparison of proxigrams with horizontal lines at 25% and 75% of normalized intensity; (Left) Comparison of microCT proxigrams, showing steeper slopes at higher resolutions indicating a sharper interface (Table 3); (Right) Comparison of LFOV nanoCT and highest resolution microCT proxigrams. The interface of the LFOV nanoCT reconstruction is much sharper (steeper) compared to the microCT volume. Further, the nanoCT interface of silica/air shows a steeper incline than silica/resin.*

*Table 3: Extracted values from the proxigrams (see also Figure S6a): The interface resolution of the reconstructions is based on the positions where the normalized intensity reaches 25 % ($x_{25\%}$) and 75 % ($x_{75\%}$) with the distance between them (Δx) indicating the interface width. The difference in normalized intensity Δy between the local extrema of the plateaus surrounding the slopes serves as a measure of the influence of phase contrast artefacts (see Table S4 for absolute intensity difference values). The slope is calculated by the quotient of 0.5/Δx describing the sharpness of the interface between silica and pores.*

| Measurement | Slope [$1/\mu m$] | $x_{25\%}$ [μm] | $x_{75\%}$ [μm] | Δx [μm] | Δy [a.u.] |
|---|---|---|---|---|---|
| MicroCT lowest resolution silica/resin | 0.39 | -0.97 | 0.34 | 1.31 | 66.1 |
| MicroCT intermediate resolution silica/resin | 0.67 | -0.38 | 0.38 | 0.76 | 54.7 |
| MicroCT highest resolution silica/resin | 0.91 | -0.1 | 0.45 | 0.46 | 47.5 |
| LFOV nanoCT silica/air | 4.32 | -0.12 | 0.00 | 0.12 | 3.7 |
| LFOV nanoCT silica/resin | 3.35 | -0.07 | 0.07 | 0.14 | 45.4 |

Having established the imaging workflows, we demonstrate how the scale-bridging tomographic data can be used to extract quantitative information about the investigated particle system. To this end, we first focus on particle size using the lowest-resolution microCT scan, as it provides the highest statistical significance. The determined particle size distribution (Figure 7a) shows a wide distribution of particles, ranging from below 10.0 µm up to 50.0 µm. The distribution exhibits a mean value of 14.5±4.7 µm. Further, the log-normal distribution fit peaks at 8.0 µm. The particle sizes are visualized in Figure 7d (see also Video S18), showing a 3D rendering of the dataset with the volume equivalent sphere diameters applied as color code.

Next, we analyzed the pore size distributions (PoSDs), using both the local thickness implementation of "XamFlow" for the lowest-resolution microCT scan and maximum sphere inscription (MSI) for the LFOV nanoCT reconstruction. The PoSD determined from the lowest-resolution microCT dataset (Figure 7b and Video S15) has a mean pore size of 6.8±2.2 µm. The log-normal distribution fit also shows a peak at 6.8 µm. The PoSD determined from the LFOV nanoCT (Figure 7c and Video S16) resolves the pores with higher resolution, ranging from below 1 µm up to 10 µm. Here, the mean of the distribution is 5.2±0.2 µm. The fitted log-normal distribution peaks at 4.3 µm. The PoSDs of both microCT and nanoCT volumes confirm the by synthesis intended pore size of 5.0 µm. This indicates that analyzing the PoSD of only one particle using LFOV nanoCT seems already representative for the entire particle fraction. The higher pore sizes of the lowest-resolution microCT dataset can be attributed to the lower resolution, leading to a loss of information on smaller pore sizes and a higher deviation. The local pore sizes in the lowest-resolution microCT reconstruction determined via the local thickness implementation of "XamFlow" and LFOV nanoCT reconstructions determined by MSI are depicted in Figure 7e (Video S15) and Figure 7f (Video S16), respectively.

Further, the overall porosity of the particles was analyzed from the lowest-resolution microCT scan. Figure S13 presents the porosity distribution for all particles, ranging from 0.0 % to 65.0 %, with a mean porosity of 22.5±13.0 %. This mean porosity is lower than the 31.4 % porosity measured for the single particle analyzed using LFOV nanoCT with a rather large diameter of 48 µm. A closer inspection of the particle size and porosity data indicates that porosity increases with particle diameter. One possible explanation is the presence of a dense, yet open, silica shell with low porosity, which has a greater impact on the overall porosity of smaller particles due to their higher surface-to-volume ratio compared to larger particles. The non-porous particles are likely fragments or particles with dimensions smaller than the pore size, potentially originating from the synthesis process or mechanical damage during particle mixing in tomography sample preparation.

Finally, we focus on the connectivity of the pore network. As evidenced from the tomography reconstructions, most of the pores are open to their particle surface and internally interconnected. Using image analysis, we determine that only a low number of closed pores is present. They can be analyzed by removing the interconnected pore system. The extracted, closed pores make up only 1.4 % of the pore volume in the LFOV nanoCT dataset. In addition, 88.9 % of these closed pores are smaller than 1.6 microns and have an average size of 0.8±0.55 µm, indicating that they are significantly below the intended pore sizes and of negligible amount in comparison to the entire pore volume.

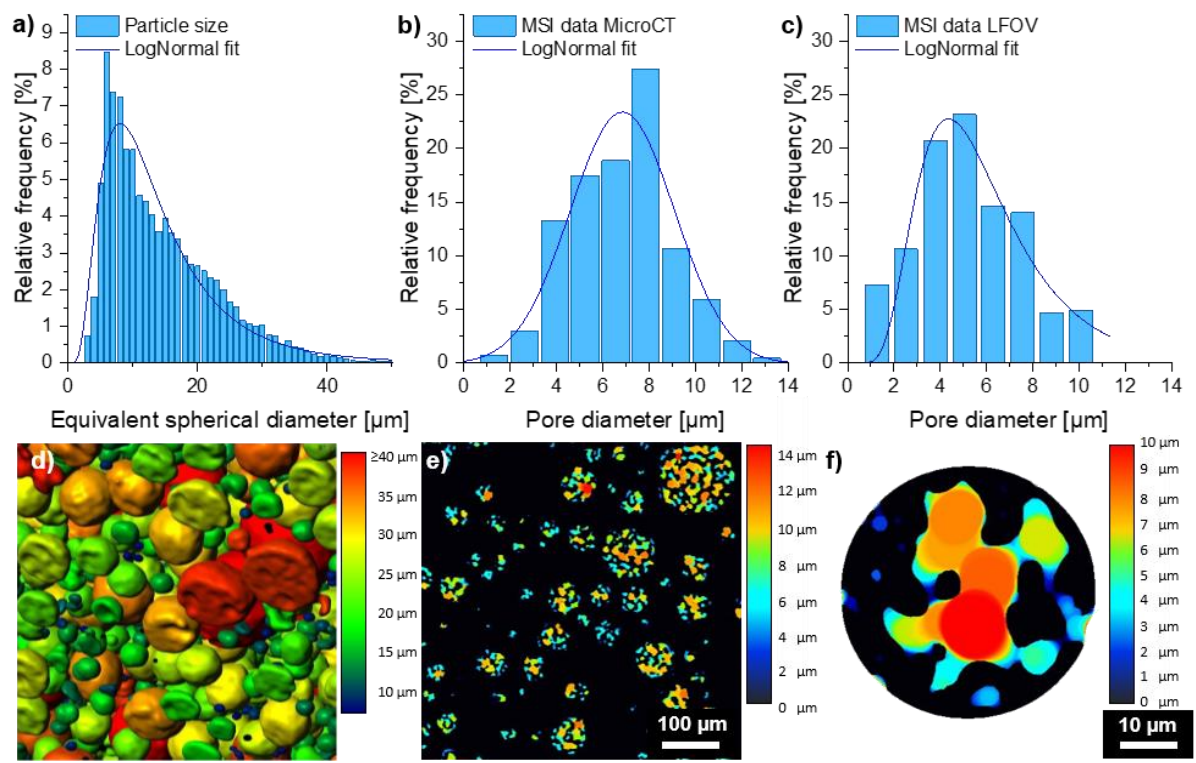

*Figure 7: a) Size distribution of the volume equivalent sphere diameter of particles measured by lowest-resolution microCT. b) Pore size distribution (PoSD) of the particles analyzed via MSI in the lowest-resolution microCT dataset. c) PoSD of the particle analyzed via MSI in the LFOV nanoCT dataset. d) 3D rendering of the particles of an excerpt of the analyzed volume from a) with equivalent diameter applied as color scale (see Video S18). e) Slice of an excerpt of the MSI of the lowest-resolution microCT dataset (see Video S15). Color code indicates range of pore diameter in µm. f) Slice of the MSI of the LFOV nanoCT dataset (see Video S16). Color code indicates range of pore diameter in µm.*

*3.6. 360° ET characterization*

The particles characterized in this work also exhibit mesopores within the walls of the macropore system.[63, 64] Such small pores cannot be resolved by the presented microCT/nanoCT approaches. Therefore, electron tomography as a complementary tomography technique is added to bridge the gap towards smallest feature sizes. To enable 360° ET, a thin pillar sample was created via FIB-SEM milling from a separately chosen, non-embedded particle. Already in single HAADF-STEM tilt series projection images (Figure 8a and Video S7), the anticipated, yet previously unseen internal mesopores are clearly discernable. These mesopores are even better resolved in the reconstructed 3D volume (Figure 8b and Video S14) forming an interconnected and open porous network towards the inner macropore surface surface. The 360° ET reconstruction has a resolution of 2.2 nm

(Table 2) determined via FSC (Figure S4f) with the ½-bit criterion. This high resolution was used to investigate the PoSD of the mesopores via MSI (Video S17). The resulting distribution ranges from below 5 nm up to 40 nm, with a mean pore size of 20.3±8.5 nm. The corresponding log-normal fit exhibits a peak at 15.6 nm (Figure 8c). Furthermore, a porosity of 31.6 % was derived for the investigated volume. As these mesopores are located throughout the whole silica material investigated in micro- and nanoCT, they constitute a non-negligible influence on the total porosity and pore size distribution of the whole system. Applying this porosity to the whole silica volume, the mean porosity of the particle of the LFOV nanoCT dataset rises from 31.4 % to 53.1 % and the mean porosity derived from the microCT data from 22.5 % to 46.8 %.

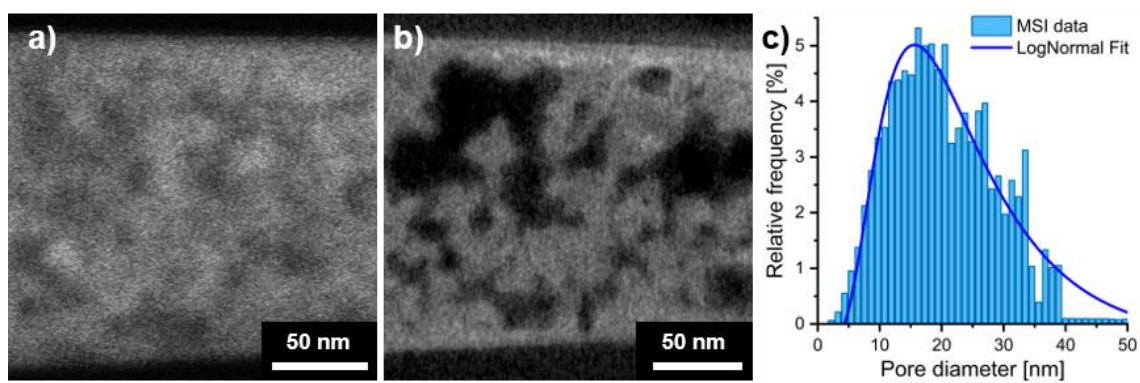

*Figure 8: High-resolution pore analysis via 360° ET of a pillar sample extracted from a separately chosen, non-embedded silica particle; a) Single projection image from the tilt series (see Video S7) in HAADF-STEM imaging mode, already revealing mesopores; b) Exemplary slice through the 3D reconstruction (see Video S14) showing clearly the interconnected mesopores present in the walls of the silica particles; c) Pore size distribution derived via MSI (see Video S17) from the reconstructed 3D volume and the corresponding log-normal fit.*

## 4. Conclusion

This study highlights the effectiveness of correlating 360° ET, nanoCT and microCT to combine high-resolution 3D information with statistically relevant volume measurements of (100 µm)³ and larger. The straightforward sample preparation workflow has proven valuable, but further tailoring of sample shapes could significantly improve imaging conditions, such as by reducing interior tomography artifacts. Techniques like laser ablation or focused ion beam (FIB) milling[39, 98, 99] could enable site-specific preparation with optimized geometries[100-102], producing refined sample tips that precisely match feature sizes and fields of view for different imaging modalities. Introducing scale-bridging markers made of high-absorbing materials and appropriate sizes could further aid 3D data correlation and fusion. These markers would provide clear contrast across imaging modalities, facilitating alignment between large field-of-view (FOV) and high-resolution techniques. Such markers could form a "3D nanoGPS" system[44-46], supporting data fusion, registration, and region of interest (ROI) identification, which are critical for advanced correlation and analysis.[3, 9, 103] 3D data registration can be performed (semi-)automatically using software solutions that apply transformations[104, 105] based on landmark-, descriptor-, or intensity-based registration approaches[106-108]. Additionally, AI-based image registration and data fusion strategies could further enhance versatility and automation.[109-112] Ultimately, improved data quality by including the above-mentioned developments could support advanced 3D data post-processing techniques, such as AI-informed segmentations[26, 39, 113] and the generation of super-resolution datasets[17, 103, 114-117]. Despite remaining a critical research area, generative AI models are expected to be pivotal in future advancements of data augmentation, annotation efficiency, and domain adaptation solutions for segmentation tasks, where the availability of annotated image data remains a critical bottleneck.[118]

One of the key strengths of the demonstrated workflow is its ability to enable local and direct analysis of pore and particle arrangements across different scales. Our study focused on optimizing the analysis of a large number of particles and their internal pores. Future research could explore the impact of different embedding materials on contrast, resolution, and segmentation precision for various particle types. Additionally, the presented method could be adapted to investigate packed particle agglomerates and their interparticle pore spaces, including larger macropores formed by interstitial sites within the packing.

By using 360° ET for mesopore analysis, we demonstrated that features below the resolution limits of both micro- and nanoCT can significantly affect material properties. Notably, the silica structure, which appeared solid in X-ray tomography, was found to have substantial intrinsic mesoporosity, significantly increasing the system's overall porosity. This highlights the potential of ET to enhance scale-bridging correlative techniques by providing precise 3D information at the nanometer scale and resolving critical features that influence material properties.

Beyond correlating the presented characterization methods of microCT, nanoCT, and ET, future studies could benefit from integrating additional 2D and 3D imaging and spectroscopy techniques, such as 3D-FIB, EDXS, and small-angle X-ray scattering (SAXS). Complementary methods for pore and particle characterization—such as analytical ultracentrifugation (AUC), dynamic light scattering (DLS), and adsorption/intrusion techniques—should also be employed for comparison and validation, further enhancing the robustness and richness of the analysis. For example, adsorption and intrusion techniques contribute valuable data for verification and enhance statistical relevance. In turn, 3D imaging techniques provide detailed local reconstructions, which can improve pore-network models and aid in interpreting intrusion and adsorption curves.

**CRediT authorship contribution statement**

**Alexander Götz:** Data curation, Formal analysis, Investigation, Methodology, Validation, Visualization, Writing – original draft, Writing – review & editing. **Alexander Kichigin:** Data curaction, Formal analysis, Investigation, Visualization, Writing – original draft, Writing – review & editing. **Johannes Böhmer:** Resources, Writing – review & editing. **Carmen Rubach:** Resources, Visualization, Writing – original draft. **Moritz Buwen:** Formal analysis, Writing – original draft, Visualization. **Benjamin Apeleo Zubiri:** Funding acquisition, Conceptualization, Project administration, Methodology, Supervision, Validation, Writing – original draft, Writing – review & editing. **Erdmann Spiecker:** Conceptualization, Funding acquisition, Methodology, Project administration, Supervision, Validation, Writing – review & editing. **Fabian Lutter:** Data curation, Formal analysis, Writing – original draft, Writing – review & editing **Sabrina Pechmann:** Data curation, Investigation. **Daniel Augsburger:** Data curation, Formal analysis, Software, Visualization. **Dennis Simon Possart:** Data curation, Formal analysis, Software, Visualization, Writing – original draft, Writing – review & editing. **Usman Arslan:** Data curation, Writing – review & editing. **Peter Suter:** Software, Writing – review & editing. **Tor Hildebrand:** Software, Data curaction, Formal analysis, Writing – original draft, Writing – review & editing. **Katharina Breininger:** Formal analysis, Software, Supervision, Writing – review & editing. **Silke Christiansen:** Conceptualization, Funding acquisition, Methodology, Project administration, Resources, Supervision, Validation, Writing – review & editing. **Umair Sultan:** Resources, Writing – original draft, Writing – review & editing. **Nicolas Vogel**: Funding acquisition, Supervision, Writing – review & editing. **Nora Vorlaufer:** Software, Writing – review & editing. **Peter Felfer:** Software, Writing – review & editing. **Matthias

**Thommes:** Resources, Data curation, Formal analysis, Funding acquisition, Writing – review & editing.

**Funding sources**

This project was funded by the Deutsche Forschungsgemeinschaft (DFG, German Research Foundation) – Project-ID 416229255 – SFB 1411.

S.P, D.P., D.A., U.A. and S.C. acknowledge the financial support from the European Union within the research project 4D + nanoSCOPE ID: 810316.

**Declaration of Competing Interest**

The authors declare that they have no known competing financial interests or personal relationships that could have appeared to influence the work reported in this paper.

**Declaration of AI tools**

During the preparation of this work the authors used OpenAI GPT-4o & DeepL Write in order to enhance the literary quality of the used phrases. After using this tool, the authors reviewed and edited the content as needed and take full responsibility for the content of the published article.

**Acknowledgements**

Thanks to Georg Haberfehlner from the Institute of Electron Microscopy and Nanoanalysis (TU Graz, Austria) for providing the "ToReAl" Matlab script for stack tilt series alignment.

**Data Availability Statement**

Data for this paper, including all data included in the figures, are available at Zenodo.org at 10.5281/zenodo.14883328.

**Glossary**

| | |
|---|---|
| 3D-FIB-SEM: | Focused ion beam scanning electron microcopy tomography |
| AMC: | Adaptive motion compensation |
| AUC: | Analytical ultracentrifugation |
| CB: | Carbon black |
| DLS: | Dynamic light scattering |
| EDXS: | Energy-dispersive X-ray spectroscopy |
| ESD: | Equivalent spherical diameter |
| ET: | Electron tomography |
| FDK: | Feldkamp-David-Kress |
| FIB: | Focused ion beam |
| FOV: | Field of view |
| FSC: | Fourier shell correlation |
| HAADF: | High-angle annular dark-field |
| HRES: | High-resolution |
| LFOV: | large-field-of-view |
| MSI: | Maximum sphere inscription |
| Macropores: | pores of > 50 nm |
| Mesopores: | pores of 2 – 50 nm |
| Micropores: | pores of < 2 nm |
| MicroCT: | Micro-computed X-ray tomography |
| NanoCT: | Nano-computed X-ray tomography |
| PEG: | Polyethylene glycol |
| PoSD: | Pore size distribution |
| ROI: | Region of interest |
| SAXS: | Small-angle X-ray scattering |
| SEM: | Scanning electron microcopy |
| SIRT: | Simultaneous iterative reconstruction technique |
| SNR: | Signal-to-noise ratio |
| STEM: | Scanning transmission electron microcopy |
| TEM: | Transmission electron microscopy |
| TEOS: | Tetraethyl orthosilicate |
| XRM: | X-ray microscope |

# Supporting information

# Correlative X-ray and electron tomography for scale-bridging, quantitative analysis of complex, hierarchical particle systems

**Alexander Götz [a], Fabian Lutter [b], Dennis Simon Possart [b,c], Daniel Augsburger [b], Usman Arslan [b], Sabrina Pechmann [b], Carmen Rubach [a], Moritz Buwen [a], Umair Sultan [d], Alexander Kichigin [a], Johannes Böhmer [a], Nora Vorlaufer [e], Peter Suter [f], Tor Hildebrand [f], Matthias Thommes [g], Peter Felfer [e], Nicolas Vogel [d], Katharina Breininger [c], Silke Christiansen [b], Benjamin Apeleo Zubiri [a,*] and Erdmann Spiecker [a]**

[a] Institute of Micro- and Nanostructure Research (IMN) & Center for Nanoanalysis and Electron Microscopy (CENEM), Friedrich-Alexander-Universität (FAU) Erlangen-Nürnberg, IZNF, Cauerstraße 3, Erlangen, Germany

[b] Fraunhofer-Institute for Ceramic Technologies and Systems – Correlative Microscopy and Materials Data, Äußere Nürnberger Straße 62, Forchheim, Germany

[c] Professur für Informatik (Pattern Recognition), Center for AI and Data Science (CAIDAS, Julius-Maximilians-Universität Würzburg, John-Skilton-Str. 4a, Würzburg, Germany

[d] Institute of Particle Technology, Friedrich-Alexander-Universität (FAU) Erlangen-Nürnberg, Cauerstraße 4, Erlangen, Germany

[e] Institute for General Materials Properties MSEI, Friedrich-Alexander-Universität (FAU) Erlangen-Nürnberg, Martensstraße 5, Erlangen, Germany

[f] Lucid Concepts AG, Zürich, Switzerland

[g] Institute of Separation Science and Technology, Friedrich-Alexander-Universität (FAU) Erlangen-Nürnberg, Egerlandstraße 3, Erlangen, Germany

* Corresponding author: benjamin.apeleo.zubiri@fau.de

# 1. Identifying regions of interest (ROIs)

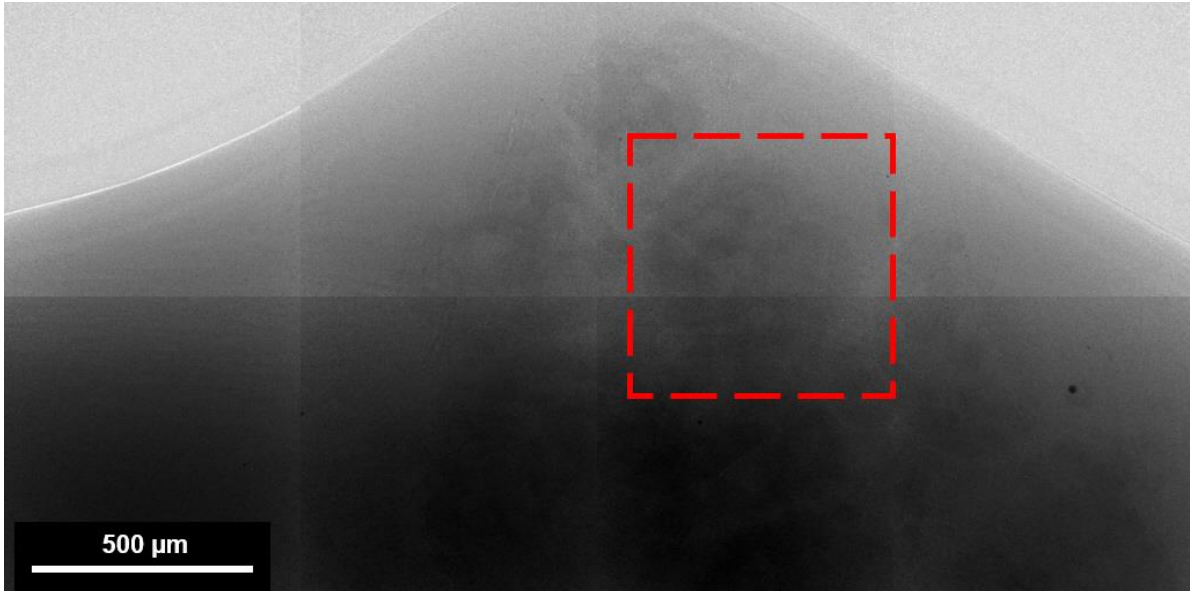

*Figure S1: By determining the position of the particle of interest in the microCT datasets in relation to features such as other particles, the tip itself or its side walls, as well as by extracting the particles size, the particle can be located in the mosaic images acquired in the Zeiss Xradia Ultra 810.*

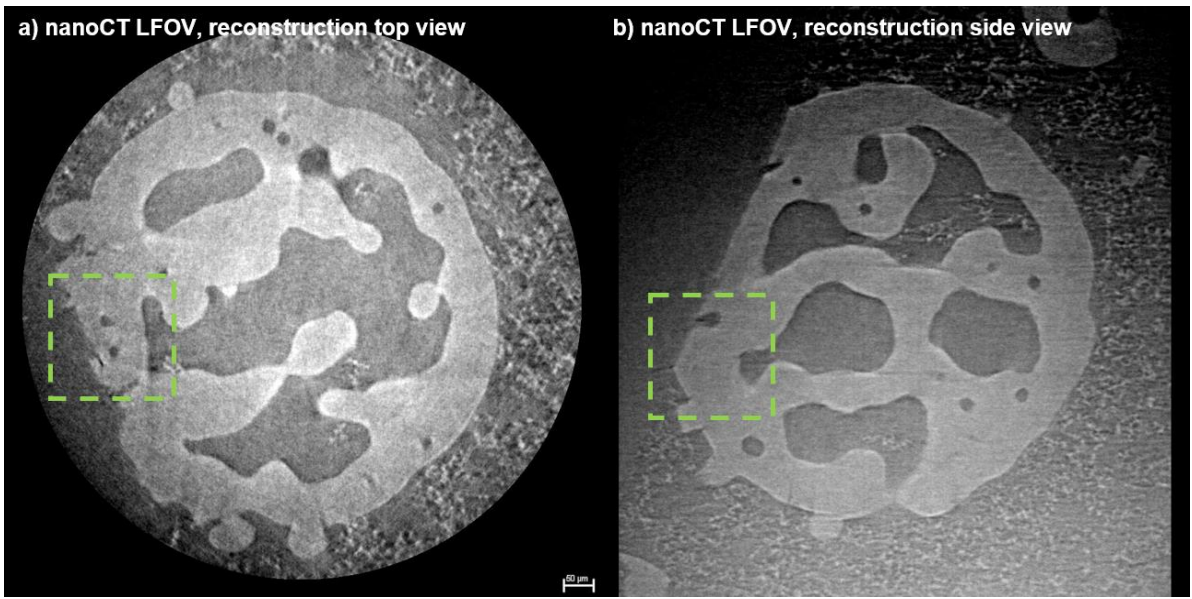

*Figure S2: Exemplary identifiers for choosing and identifying the region of interest (ROI) for the HRES nanoCT scan in a LFOV nanoCT reconstruction. a) A pore constellation and a fracture were chosen as ROI, marked in green, in a slice through the LFOV nanoCT reconstruction; b) Finding the center of the constellation in the side view (slice through the LFOV nanoCT reconstruction) is important to fully correlate the position of the features relative to the particle. By calculating the distance of the ROI center from the top, front and side of the particle, the ROI can be located in the transmission image in HRES nanoCT mode in front and side view, allowing targeting of the desired ROI for the HRES nanoCT scan.*

## 2. Tilt series

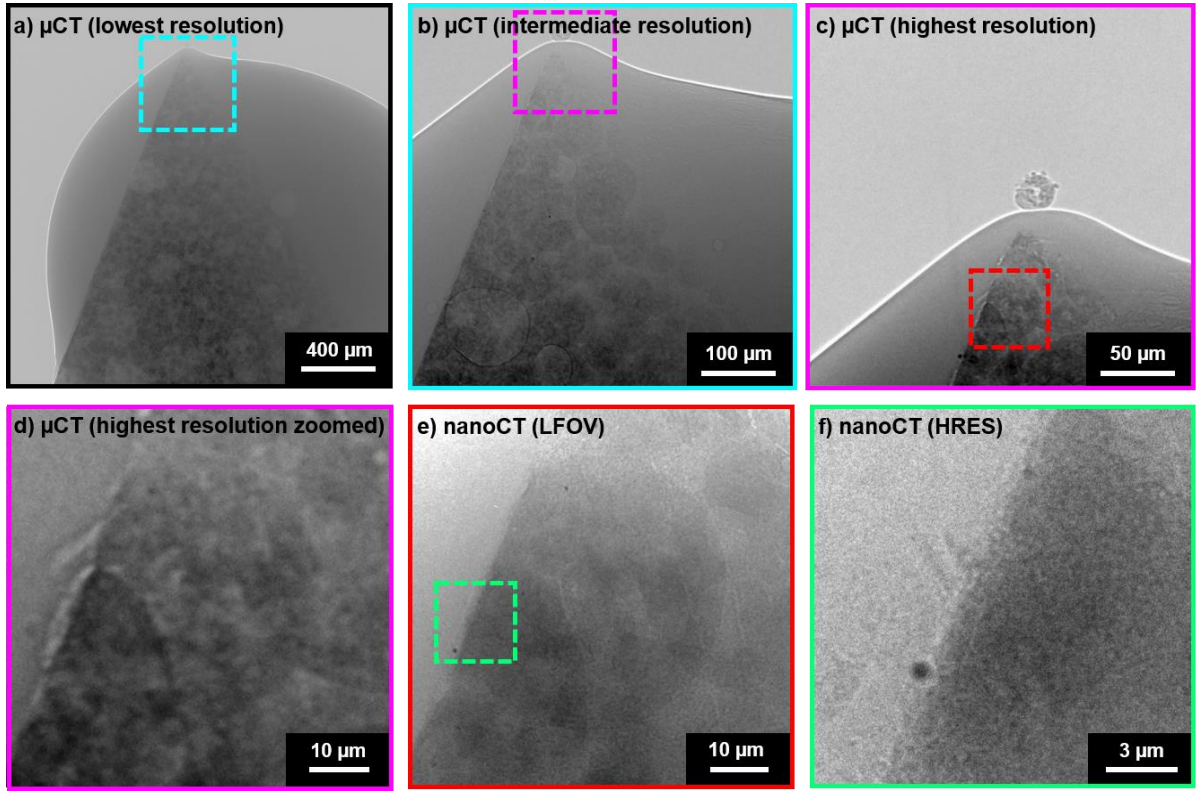

*Figure S3: Projections from all X-ray CT tilt series acquired (cf. Figure 4) and the next smaller resolution mode marked as a colored box: a) Lowest resolution microCT projection showing more than 14,000 particles in dark contrast with multiple residual air bubbles in medium grey in between and a stronger contrast at the interface. The surrounding resin matrix and glue are visible in dark grey and the background in light grey; b) Intermediate resolution microCT projection with the same contrasts as a) but showing less particles and only few air bubbles; c) Highest resolution microCT projection showing even less particles but in a higher resolution and no air bubbles, with particles in black to dark grey, resin and glue in medium grey and background in black; d) The highest resolution microCT projection cropped and zoomed to the particle visible in e), which shows a projection from the LFOV nanoCT tilt series with the particles in dark to medium grey, depending on size and distance from the sampled volume of (64 µm)³. This projection is visibly higher resolved and seems to have a better SNR than d); f) Projection from the HRES nanoCT tiltseries showing the ROI in side view. Here, the interface between the particle and the background is clearly visible but due to the projection being a transmission image, the particle structure is hardly discernable.*

Table S1: List of videos supplied for this publication. The analysis videos contain the pore size analysis in form of the local thickness or MSI, and for the low resolution µCT dataset, also the particle rendering with the particle size applied as color scale.

| Imaging mode | Tilt series video | Reconstruction video | Analysis video |
|---|---|---|---|
| **Low resolution µCT (4x)** | Video S1 | Video S8 | Video S15 & Video S18 |
| **Medium resolution µCT (20x)** | Video S2 | Video S9 | - |
| **High resolution µCT (40x)** | Video S3 | Video S10 | - |
| **LFOV nanoCT resin** | Video S4 | Video S11 | Video S16 |
| **LFOV nanoCT air** | Video S5 | Video S12 | - |
| **HRES nanoCT** | Video S6 | Video S13 | - |
| **Electron tomography** | Video S7 | Video S14 | Video S17 |

## 3. Overview of 3D data analysis methods

Table S2: Overview over the applied 3D data analysis methods.

| Imaging mode | Registration transformation via arivis | Rotation axis | Segmentation software | Pore size analysis software | Particle size analysis software | Resolution analysis | Contrast analysis |
|---|---|---|---|---|---|---|---|
| **Low resolution µCT (4x)** | Rotation | z & y | | ImageJ MSI | XamFlow | FSC | Proxigram |
| **Medium resolution µCT (20x)** | Rotation | z & y | - | - | arivis | FSC | Proxigram |
| **High resolution µCT (40x)** | Rotation | z & y | - | - | arivis | FSC | Proxigram |
| **LFOV nanoCT** | Registration source | – | arivis | ImageJ MSI | arivis | FSC | Proxigram |
| **HRES nanoCT** | Rotation | z | - | – | – | FSC | - |
| **Electron tomography** | - | - | ImageJ Threshold | ImageJ MSI | - | FSC | - |

# 4. Fourier Shell Correlation (FSC)

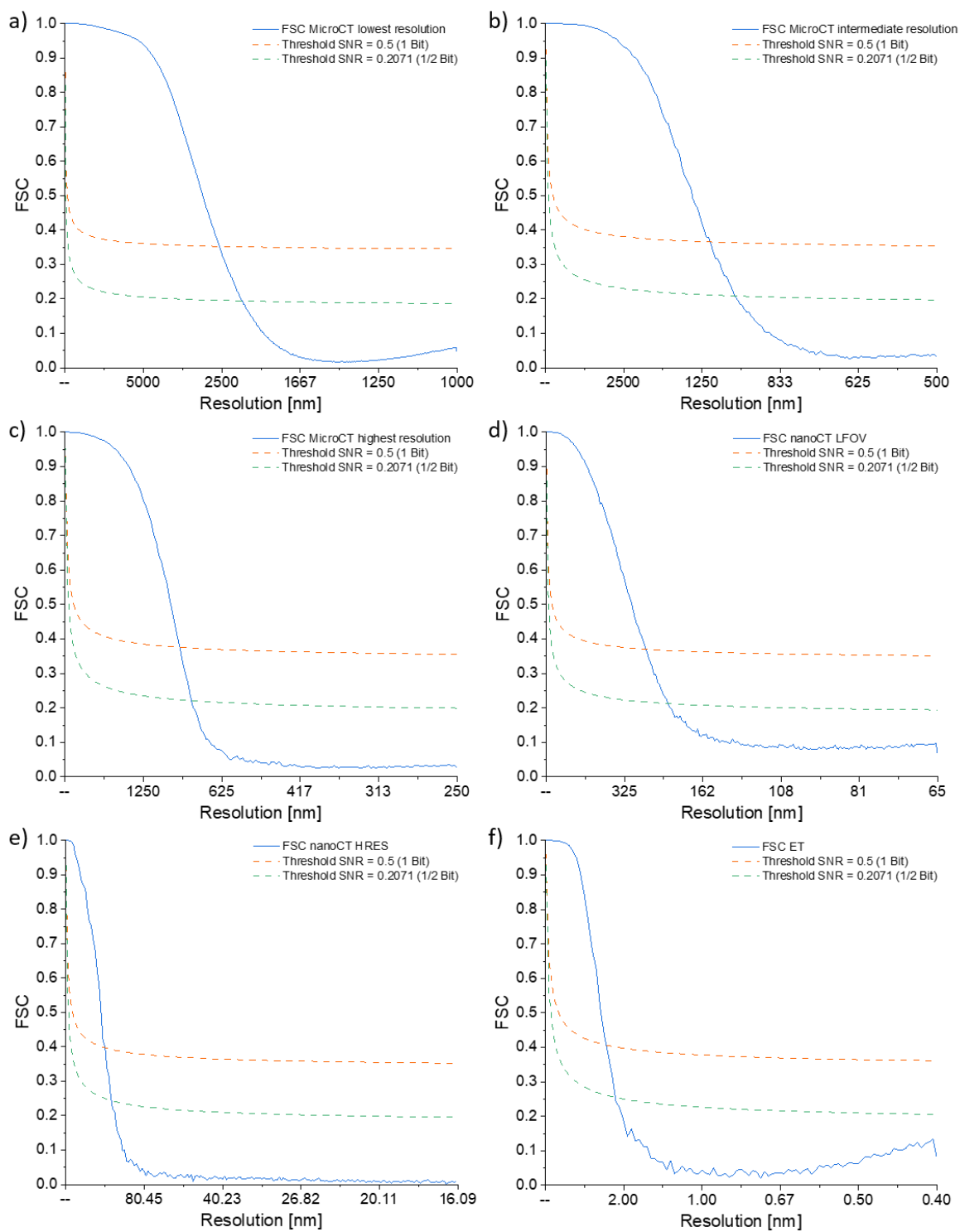

*Figure S4: Fourier Shell Correlations (FSCs) for all datasets acquired for this work. The FSC data is plotted against the resolution. The intersections of the curves with the threshold curve of either the 1 bit or ½-bit criterion can be used as a measure for the resolution.*

*Table S3: List of resolutions determined via Fourier-Shell-Correlation (FSC) for all datasets (Figure S4). The 1-bit criterion leads to a more conservative resolution determination, whereas the ½-bit criterion is well established as an accurate way to determine precise resolutions*

| Imaging mode | ½-bit criterion resolution [nm] | 1-bit criterion resolution [nm] |
|---|---|---|
| **Low resolution µCT (4x)** | 2207 | 2538 |
| **Medium resolution µCT (20x)** | 1031 | 1182 |
| **High resolution µCT (40x)** | 774 | 852 |
| **LFOV nanoCT** | 208 | 257 |
| **HRES nanoCT** | 138 | 160 |
| **Electron tomography** | 2.2 | 2.6 |

# 5. Proxigrams

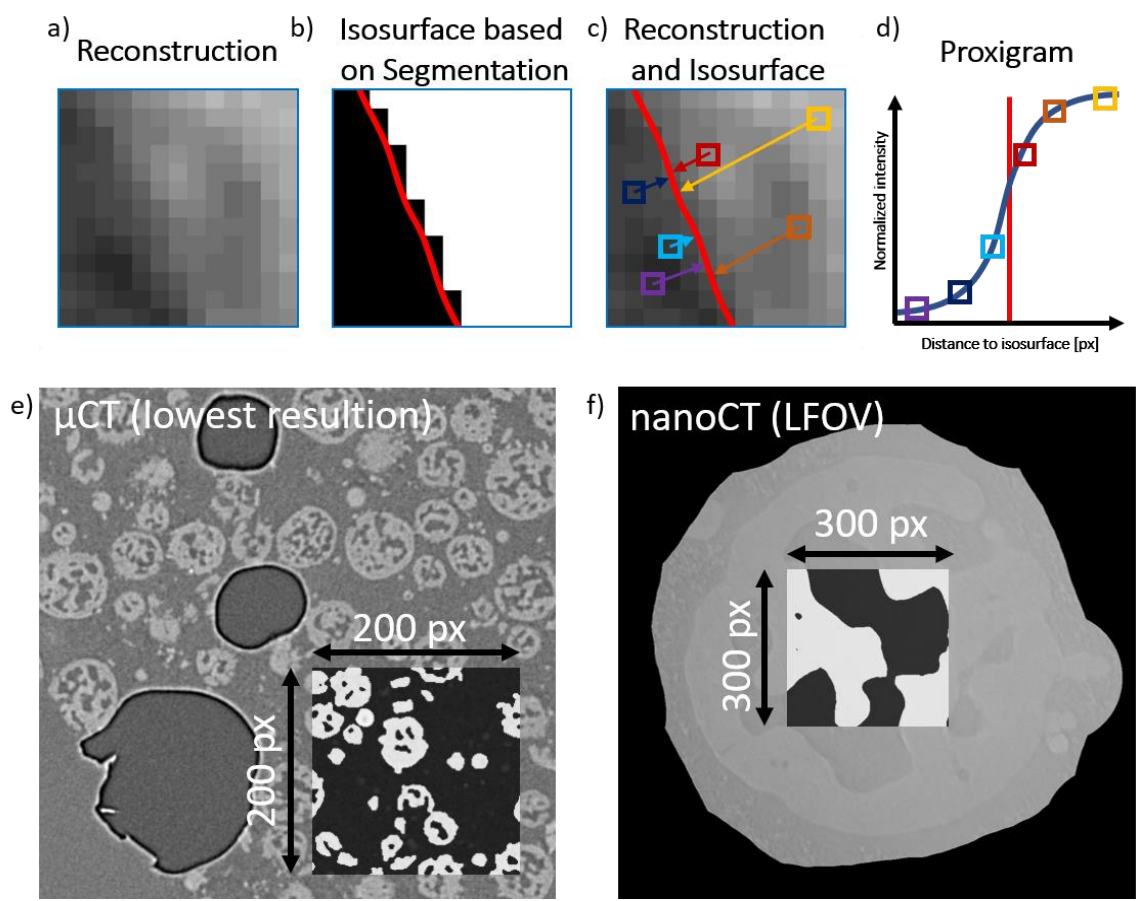

*Figure S5: Scheme to illustrate the generation of a proxigram. a) The reconstruction consists of two phases, solid and pore space; b) The segmentation is a binary representation of the pore and solid space present in the sample. It is used to define*

*an isosurface along the interface, which is then applied to the selected volume of the 3D reconstruction; c) The isosurface is used in the reconstruction to generate intensity profiles perpendicular to the isosurface and therefore perpendicular to the material interface; d) The intensity profiles are averaged and normalized with the lowest average intensity registered as 0 and the maximum average intensity registered as 1. Then, the normalized intensity is plotted against the distance to the isosurface to generate the proxigram. e) & f) The volume utilized for a proxigram is exemplary visualized by overlaying the reconstruction slice (greyscale image) at the center of utilized region with the corresponding segmentation slice (binarized image) cropped to the size of the proxigram region. Typically, this region is a cube of 300x300x300 voxels, however, only the desired interface should be present in this volume. Therefore, the size of the volume utilized for the proxigram can vary. The microCT datasets utilize only a volume of 200x200x200 voxels to avoid incorporating bubbles and volume outside of the sample, whereas the LFOV nanoCT datasets utilize a volume of 300x300x300 voxels.*

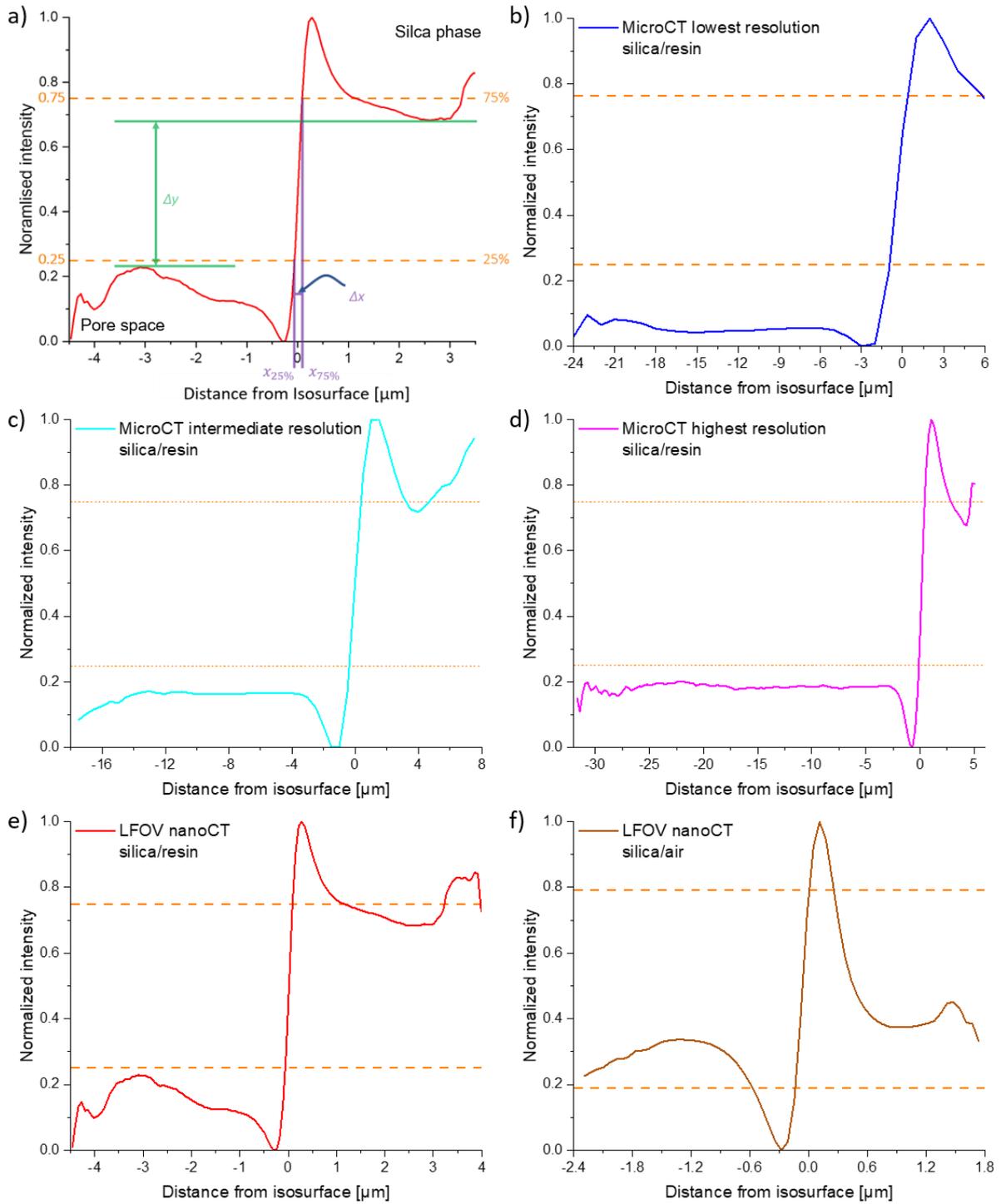

*Figure S6: a) Exemplary proxigram visualizing the extracted values from the proxigrams. The vertical, orange lines are at 75 % and 25 % (respectively the values of 0.25 and 0.75) on the y-axis. The markings in green indicate the distance Δy and the violet markings indicate the values $x_{25\%}$, $x_{75\%}$ and the distance Δx. The slope of the proxigram between the $x_{25\%}$ $x_{75\%}$ is determined as a measure for the sharpness of the interface. b-f) Proxigrams of the different microCT and LFOV nanoCT scans.*

*Table S4: Comparison of Δy in [%] and in [counts] to emphasize the difference between the phase contrast artefacts.*

| Dataset analysed | Δy [%] | Δy [counts] |
|---|---|---|
| **nanoCT LFOV silica/air** | 3.7 | 191 |
| **nanoCT LFOV silica/resin** | 45.4 | 1140 |
| **MicroCT highest resolution silica/resin** | 47.5 | 1061 |
| **MicroCT intermediate resolution silica/resin** | 54.7 | 421 |
| **MicroCT lowest resolution silica/resin** | 66.1 | 1093 |

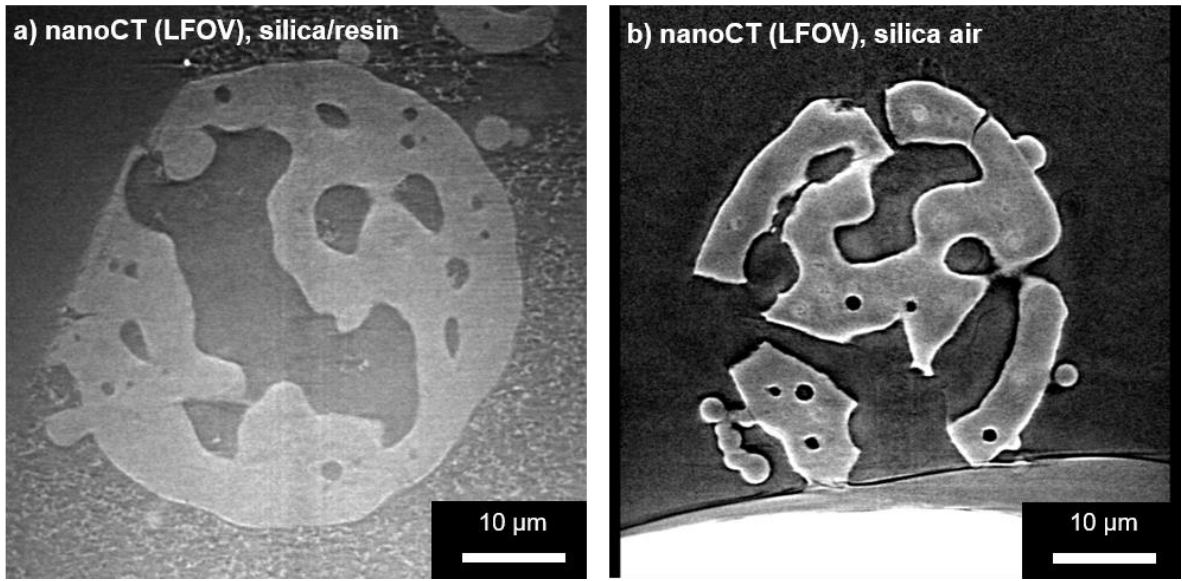

*Figure S7: Visualization of the varying phase contrast enhancement in different media; a) Silica particle embedded in resin (Videos S4 and S11); b) Silica particle in air (Videos S5 and S12). The contrast of the silica in a) looks far more consistant than in b), as well as more distinct from the background. Further, the halo artefacts are for more intense in b) than in a).*

# 6. Workflow XamFlow

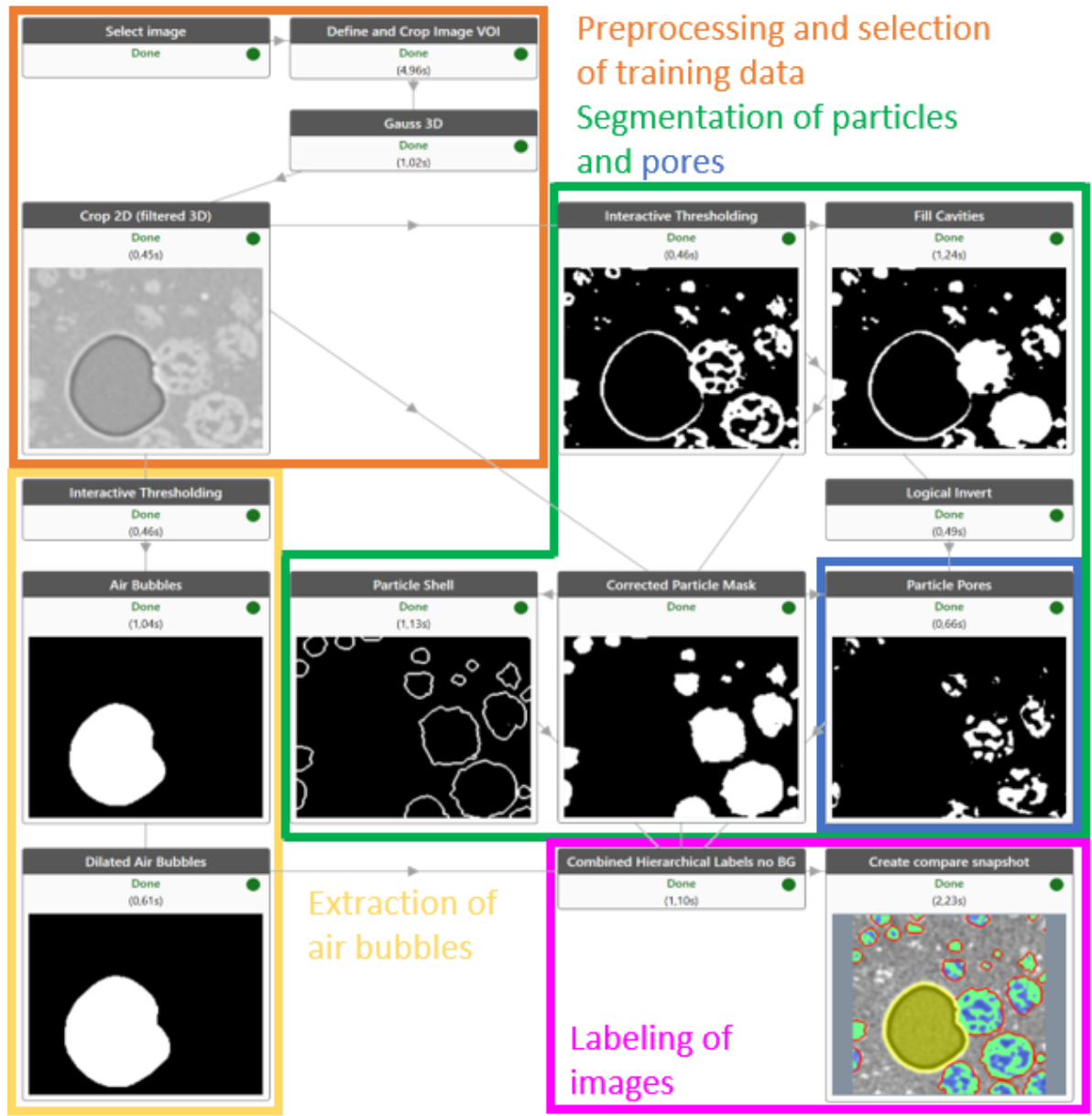

*Figure S8: First workflow step (1) of the XamFlow analysis, the generation of the 2D training data for the Unet segmentation. It can be split into four sections: The preprocessing and selection of training data, the segmentation of particles and pores, the extraction of air bubbles and finally the labeling of the images.*

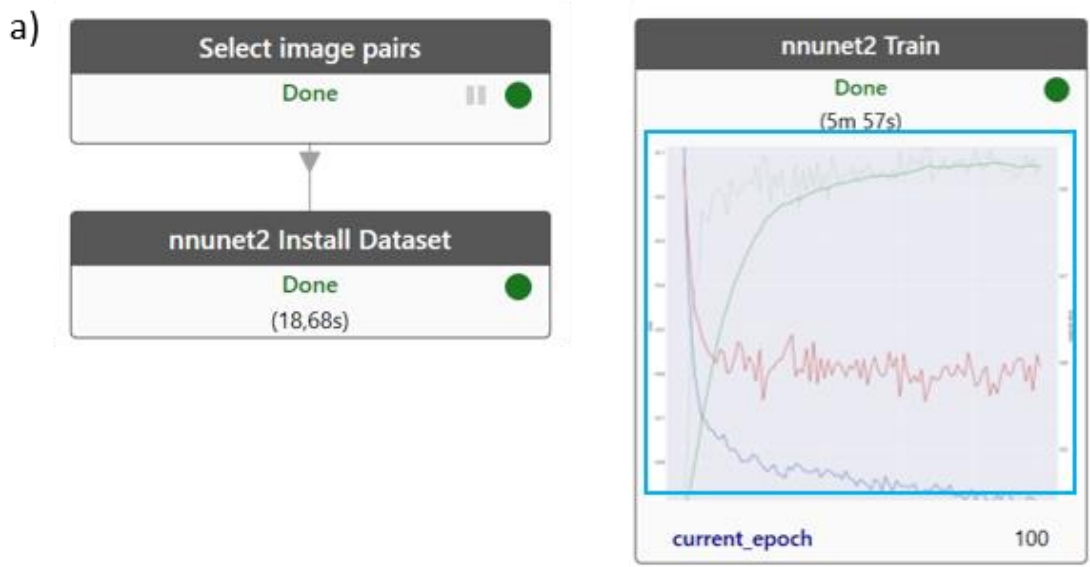
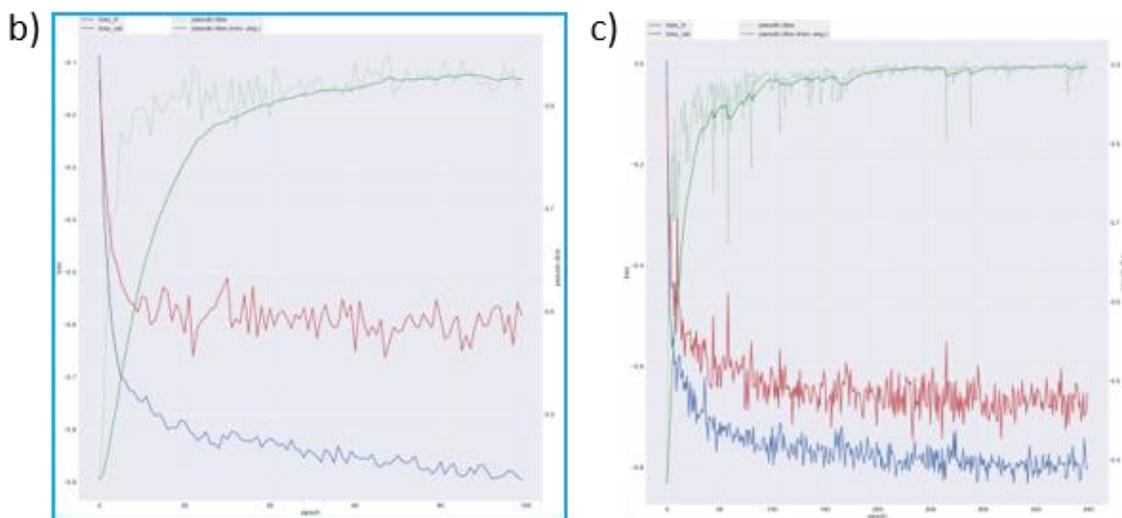

Figure S9: a) The training of a 2D Unet model, trained at the end of workflow step one (1) Training graph of the 2D Unet model of workflow step one (1); b) ; c) Training graph of the 3D Unet model of workflow step two (2).

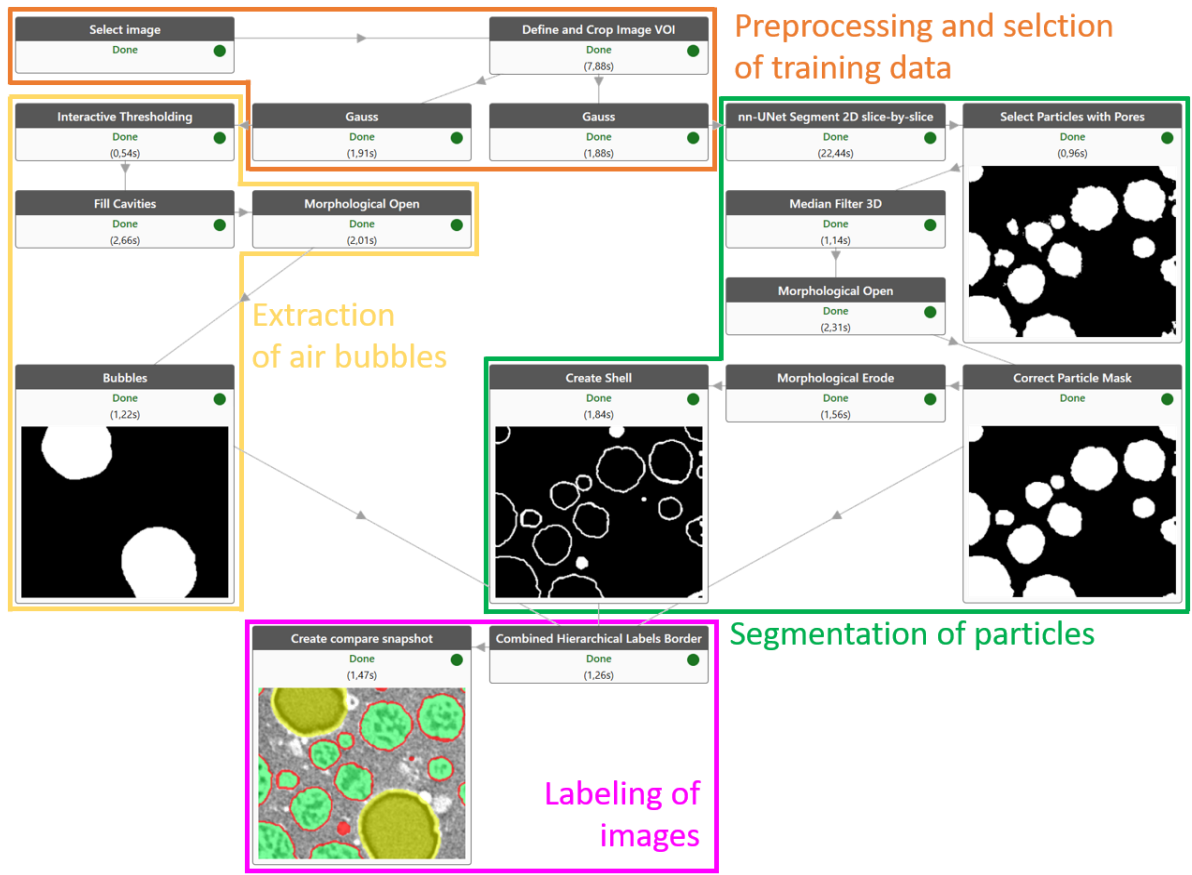

*Figure S10: Workflow step two (2), the generation of 3D training data for the Unet segmentation. It can be separated into the same four parts as the first workflow step (1): The preprocessing and selection of training data, the segmentation of particles, this time without the separation of the pores, the extraction of air bubbles and finally the labeling of the images.*

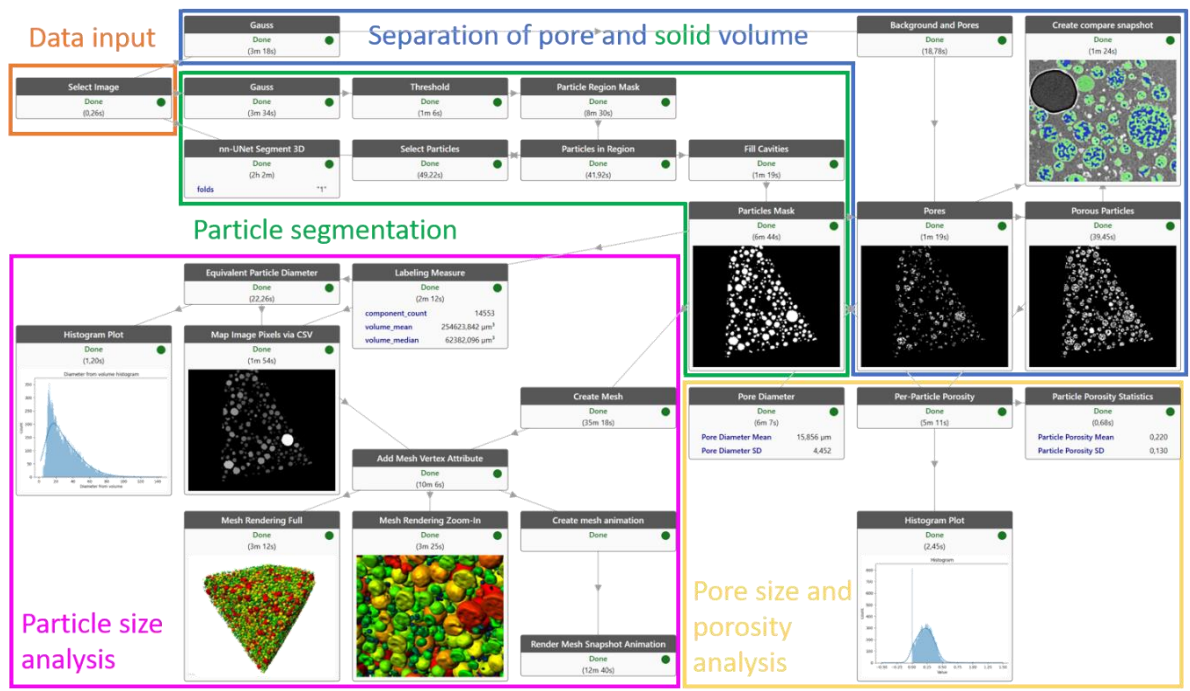

*Figure S11: The third and final workflow step (3). It can be separated into five parts: The input of the data, the segmentation of the particles, the particle size analysis based on this segmentation, the separation of the particles into solid and pore volume and the pore size and porosity analysis based on the separated data.*

# 8. Estimation of resolution in microCT

To determine the resolution in the micro-CT measurements of this work, we used the full-width-half-maximum (FWHM) of the derivate of the line profile at the edge of one embedded, phase separated particle. By this, we can estimate the line spread and the corresponding resolution. As an example, one chosen line profile, its derivate and the corresponding Gaussian peak fit, is shown in Figure S12 for the intermediate resolution scan, with a pixel size of 500 nm. This was done for all three microCT scans ten times at different positions in different xz-slices (perpendicular to the rotation axis). Afterwards the mean value and its error were calculated. The results are shown in Table S5.

*Table S5: Parameters from the resolution estimation analysis for the microCT experiments, comprising the image pixel size, the line spread of the line scans and the estimated resolutions (see Figure S12).*

| Dataset name | Pixel size [nm] | Line spread [px] | Estimated resolution [µm] |
|---|---|---|---|
| **Highest resolution microCT** | 250 | 3.71 ± 0.25 | 0.928 ± 0.063 |
| **Intermediate resolution microCT** | 500 | 3.04 ± 0.18 | 1.520 ± 0.092 |
| **Lowest resolution microCT** | 1000 | 2.52 ± 0.13 | 2.52 ± 0.13 |

The resolution is improving as the sampling becomes finer. This is to be expected as the resolution limit by the manufacturer is given as 500 nm (at 40 nm sampling/pixel size) in 2D [1]. This is also indicated by the increasing line spread, when the resolution limit comes closer.

The resolution that is estimated in this way, is an in-plane resolution, as all line profiles are in the xz-plane. The off-plane (out-of-plane) resolution, without a preferred direction of the line profiles, is usually inferior to the in-plane resolution. This is due to the necessary interpolation between the voxels along the line to generate the line profile.

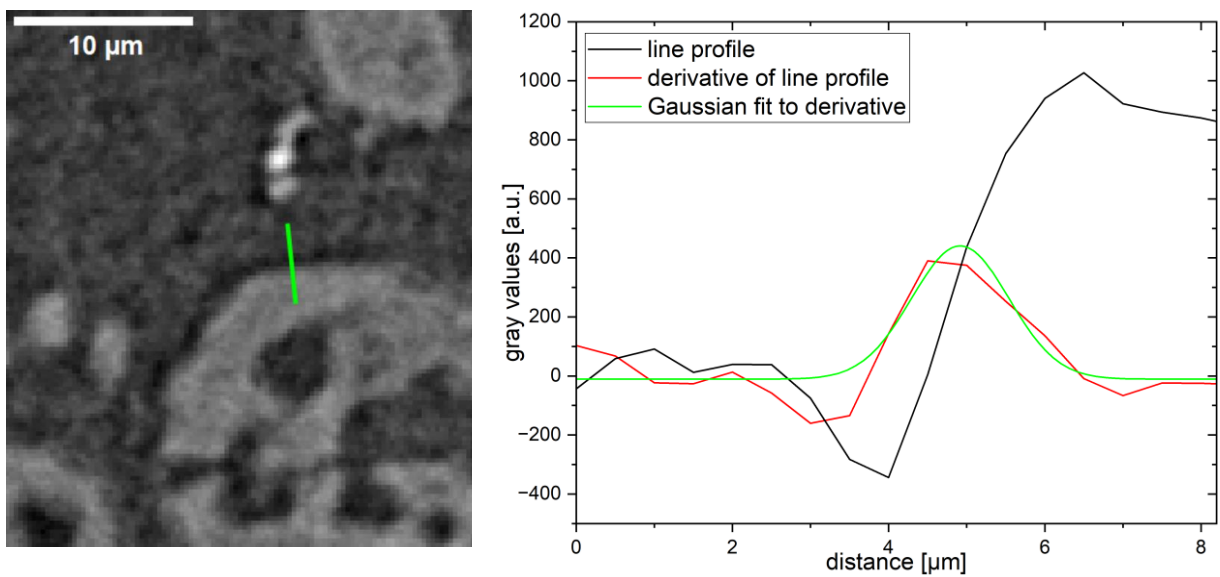

*Figure S12: Exemplary routine for the resolution estimation for microCT 3D reconstructions (see Table 2, value "Resolution estimated"), showcased for the intermediate resolution (20x) microCT scan at 500 nm sampling (pixel size); (left) xt-Slice through the reconstructed volume, where the evaluated line profile at the edge of one particle is exemplarily*

*marked in green. (Right) Plot of the corresponding line profile, its derivative and the Gaussian peak fit to estimate the line spread and the corresponding resolution. In this example, the resolution is at 1.47 μm (full width at half maximum, FWHM) corresponding to a line spread of 2.94 px.*

## 9. Porosity distribution

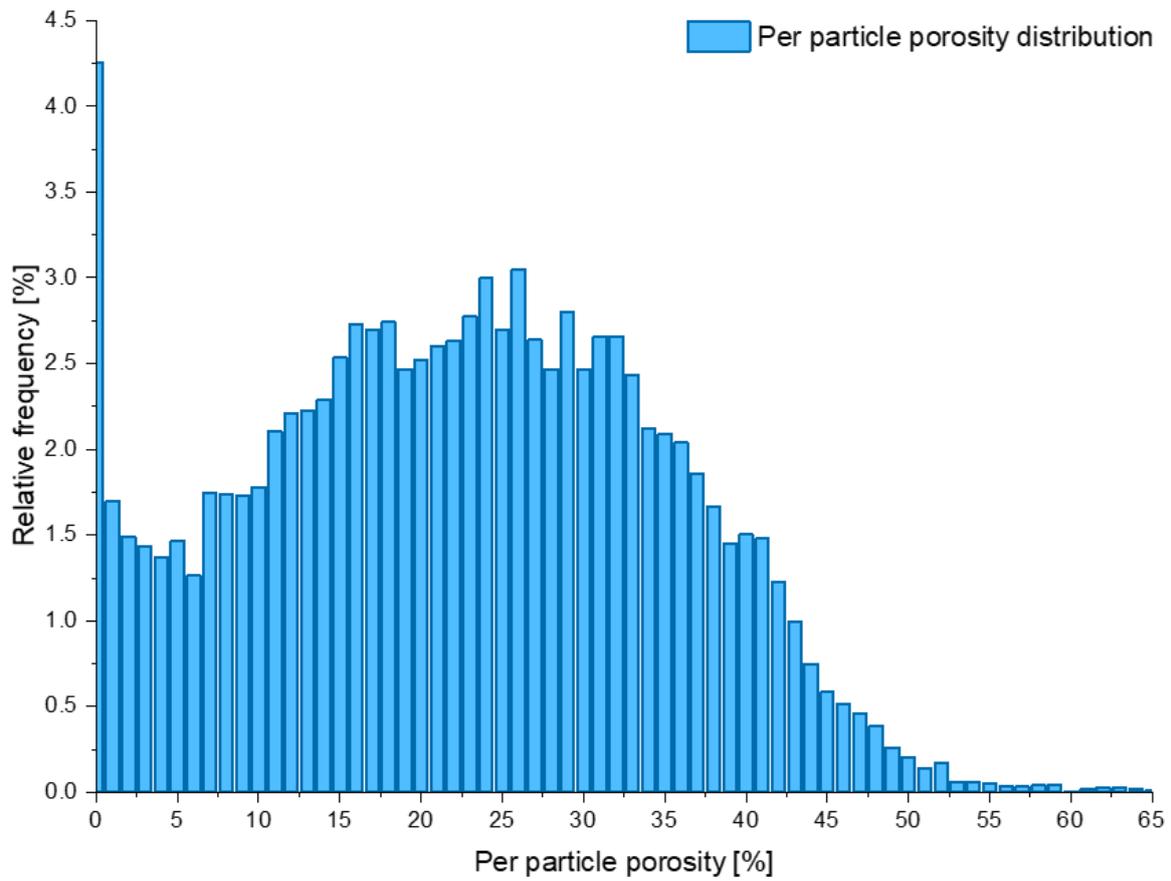

*Figure S13: Per particle porosity distribution of the lowest-resolution microCT dataset with an average porosity of 22.5 ± 13.0 %. At lower porosities, a rise in relative frequency is present, with a peak to 0 % porosity. This is probably due to particles below the intended pore size of 5 μm and particle fragments.*

## 10. Particle size rendering

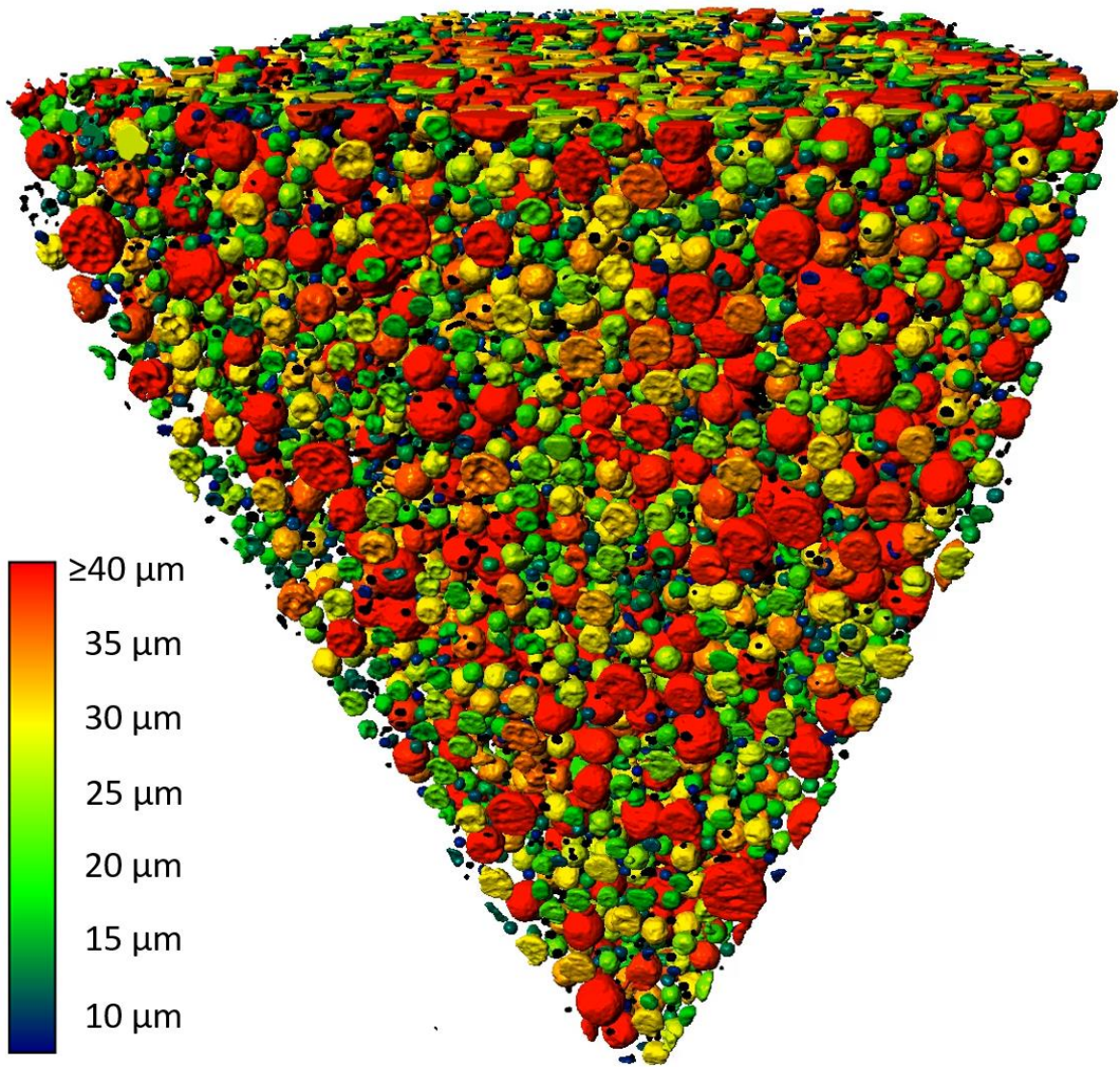

*Figure S14: Rendering of the particles derived from the whole lowest-resolution microCT dataset. The particle size was applied as a color scale. A video of this rendering can be found in Video S18.*

## 11. Influence of interior tomography/sample size on transmission intensity

Since the sample dimensions exceed the field of view (FOV) of the nanoCT measurements (Figure 5e,f), we analyzed how the measured transmission intensity depends on the local sample size. In this analysis, the local sample composition and thickness along multiple lines—indicated in Figure S15—in the corresponding region of the lowest-resolution microCT reconstruction were estimated, and the corresponding transmission intensities were calculated (see Table S6). All lines are intersecting the particle investigated with nanoCT. Notably, silica particles outside the nanoCT reconstruction's FOV significantly reduce the measured intensity, while the influence of glue and resin is considerably weaker.

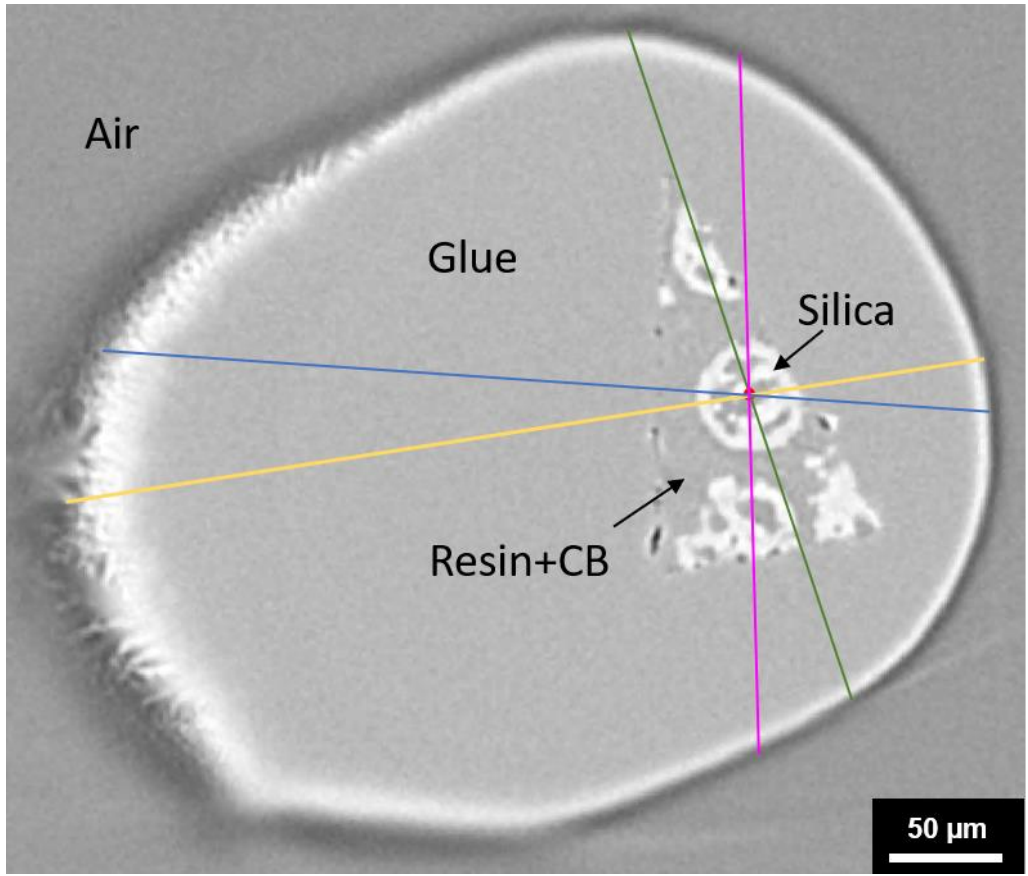

*Figure S155: Slice through the lowest-resolution microCT reconstruction perpendicular to the long axis of the pillar sample at the position of the nanoCT scans from Figure 5e,f (as well as Figure S2b, Figure S3e, Figure S7a). The lines indicate where the local total thickness and constituent materials were extracted for transmission intensity estimation for nanoCT in Table 6. All lines are intersecting the particle investigated with nanoCT.*

*Table S6: Dependence of measured transmission intensity on the local sample size exceeding the FOV in nanoCT (Figure 5e,f). In this analysis, the local sample composition and thickness along multiple lines, as indicated in Figure S15, in the corresponding region of the lowest-resolution microCT reconstruction were estimated, and the corresponding transmission intensities were calculated. Especially silica particles outside the FOV of the nanoCT reconstructions strongly lower the measured intensity, whereas the influence of glue and resin is far weaker.*

| Line in Figure S15 | Thickness [µm] of | | | | Total | Intensity transmission [%] |
| --- | --- | --- | --- | --- | --- | --- |
| | Silica in ROI of LFOV nanoCT | Resin & CB | Glue | Surrounding silica | | |
| **Yellow** | 27 | 42 | 336 | 0 | 405 | 7.3 |
| **Green** | 34 | 92 | 138 | 50 | 314 | 3.7 |
| **Magenta** | 33 | 74 | 171 | 31 | 309 | 5.7 |
| **Blue** | 29 | 42 | 323 | 0 | 394 | 7.5 |

## Data Availability Statement

Data for this paper, including all data included in the figures, are available at Zenodo.org at 10.5281/zenodo.14883328.